\documentclass[nofootinbib,english,preprint,amsmath,amssymb,aps]{revtex4-1}
\usepackage{amsmath}
\usepackage{graphicx}
\usepackage{esint}
\usepackage{subfig}
\usepackage[singlelinecheck=false,justification=raggedright]{caption}
\providecommand{\tabularnewline}{\\}
\makeatletter
\@ifundefined{textcolor}{}
{%
 \definecolor{BLACK}{gray}{0}
 \definecolor{WHITE}{gray}{1}
 \definecolor{RED}{rgb}{1,0,0}
 \definecolor{GREEN}{rgb}{0,1,0}
 \definecolor{BLUE}{rgb}{0,0,1}
 \definecolor{CYAN}{cmyk}{1,0,0,0}
 \definecolor{MAGENTA}{cmyk}{0,1,0,0}
 \definecolor{YELLOW}{cmyk}{0,0,1,0}
}

\usepackage{bm}% bold math
\usepackage[mathlines]{lineno}% Enable numbering of text and display math
\usepackage[english]{babel}
\usepackage[autostyle, english = american]{csquotes}
\usepackage{babel}

\begin{document} 

%\begin{refsection}

\title{Molecular cavity optomechanics: a theory of plasmon-enhanced Raman scattering}

\author{Philippe Roelli$^{1}$, Christophe Galland$^{1}$, Nicolas Piro$^{1}$, Tobias J. Kippenberg$^{1\ast}$}

\affiliation{$^{1}$\'Ecole Polytechnique F\'ed\'erale de Lausanne (EPFL), CH-1015 Lausanne,
Switzerland }

\date{\today}

%%%%%%%%%%%%%%%%% END OF PREAMBLE %%%%%%%%%%%%%%%%

\begin{abstract}

  \textbf{%The exceptional enhancement of Raman scattering cross-section by localized plasmonic resonances in the near-field of metallic surfaces, nanoparticles or tips has enabled spectroscopic fingerprinting of single-molecules and is widely used in material, chemical and biomedical analysis. We present a new theory of plasmon-enhanced Raman scattering by mapping the problem onto the canonical model of cavity optomechanics, in which a bi-directional interaction takes place between molecular vibration and plasmon. The optomechanical coupling rate, from which we derive the Raman cross section, is computed from the molecule’s Raman activities and the plasmonic field distribution. When the excitation is blue-detuned from the plasmon onto the vibrational sideband, the electromagnetic force can lead to dynamical backaction amplification of molecular vibrations, revealing an enhancement mechanism not contemplated before. The optomechanical theory provides a quantitative framework for the calculation of enhanced cross-sections, recovers known results, and enables the design of novel systems that leverage dynamical backaction to achieve additional, mode-selective enhancement. It yields a new understanding of plasmon-enhanced Raman scattering and opens a route to molecular quantum optomechanics.
  The exceptional enhancement of Raman scattering cross-section by localized plasmonic resonances in the near-field of metallic surfaces, nanoparticles or tips has enabled spectroscopic fingerprinting of single molecules and is widely used in material, chemical and biomedical analysis. {The conventional explanation attributes the enhancement to the antenna effect focusing the electromagnetic field into sub-wavelength volumes. Here we introduce a new model that additionally accounts for the dynamical and coherent nature of the plasmon-molecule interaction and thereby reveals an enhancement mechanism not contemplated before: dynamical backaction amplification of molecular vibrations.} We first map the problem onto the canonical model of cavity optomechanics, in which the molecular vibration and the plasmon are {parametrically} coupled. The optomechanical coupling rate, from which we derive the Raman cross section, is computed from the molecules Raman activities and the plasmonic field distribution. {When the plasmon decay rate is comparable or smaller than the vibrational frequency} and the excitation laser is blue-detuned from the plasmon onto the vibrational sideband, the resulting delayed feedback force can lead to {efficient parametric} amplification of molecular vibrations. The optomechanical theory provides a quantitative framework for the calculation of enhanced cross-sections, recovers known results, and enables the design of novel systems that leverage dynamical backaction to achieve additional, mode-selective enhancement. It yields a new understanding of plasmon-enhanced Raman scattering and opens a route to molecular quantum optomechanics.} 
  
\end{abstract}
\maketitle 
%\onecolumngrid

\section*{Introduction}

In 1973 Fleischmann \textit{et al.} first reported the dramatic enhancement of the Raman scattering cross-section of molecules on rough metal surfaces \cite{Fleischmann_1974}, an effect confirmed in 1977 by Van Duyne   \textit{et al.} \cite{VanDuyne1977}. Two decades later, this technique known as surface-enhanced Raman scattering (SERS) enabled the detection of single molecules \cite{Kneipp1997,NieScience}. Enhancement
factors in the range of $10^{10}$ - $10^{14}$ %though some results are disputed \cite{Etchegoin}. 
have been reported to occur at ``hot spots'' \cite{Shalaev}, regions of high electromagnetic fields associated with localized plasmonic resonances. Using the plasmon at the tip of a scanning tunneling microscope \cite{Pettinger} has led to a powerful analytical tool for sensitive Raman imaging: TERS (tip-enhanced Raman scattering). SERS and TERS both rely on the phenomenon of plasmon-enhanced Raman scattering, which is today widely employed in the fields of material and surface science \cite{Pettinger}, nanotechnology \cite{Sharma}, chemistry \cite{Kneipp1997} and even in-vivo biomedical applications \cite{Qian}. 

The generally accepted model for SERS invokes the combined enhancement by the plasmonic hot spot of the incoming electromagnetic field and the Raman scattered field \cite{Shalaev,Pendry}. It predicts an enhancement of the Raman cross-section proportional to the fourth power of the field enhancement. Although this ``$E^{4}$ law" has been verified experimentally, the observation of even larger enhancements and of anomalous Stokes/anti-Stokes intensity ratio \cite{Maher2004} have raised the suspicion that a ``vibrational pumping" mechanism was involved \cite{LeRu2006b}. Moreover in recent experiments the maximal enhancement was achieved when the laser was blue-detuned from the plasmon resonance by the vibrational frequency (exciting the anti-Stokes vibrational sideband) \cite{Zhang2013,Zhu_2014}. Under these conditions large nonlinear effects were also evidenced in \cite{Zhang2013,Jiang_2015}. These observations are calling for a new theoretical understanding and for further investigations \cite{atkin}. 

\begin{figure}
\begin{centering}
\includegraphics[scale=0.55]
{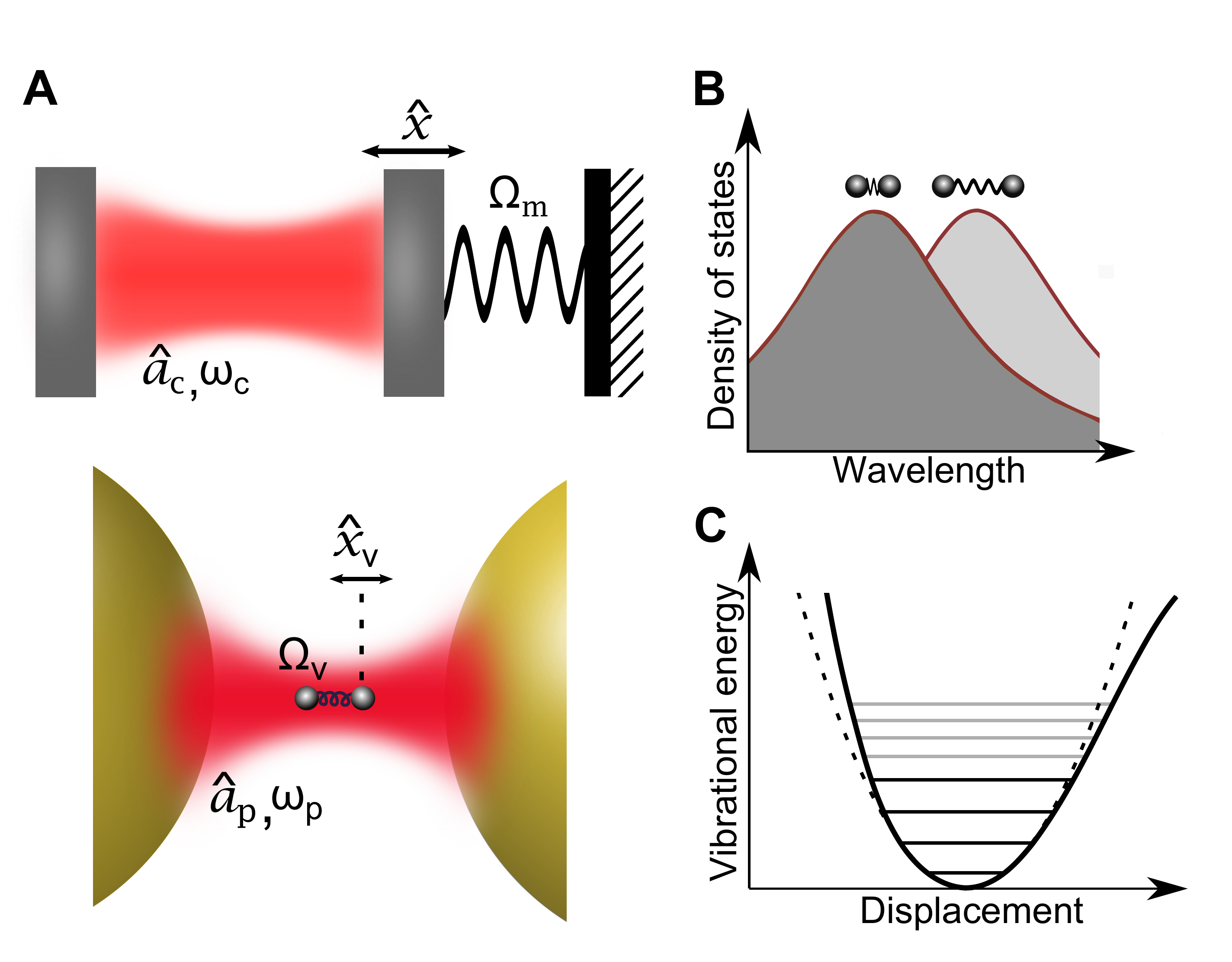}
\par\end{centering}

\protect\caption{\label{fig:Intro}
\textbf{Cavity-optomechanical model of the interaction between plasmon and molecular vibration.}
(\textbf{A}) Schematic mapping between (upper panel) an optical cavity with a mechanically compliant mirror and (lower panel) a plasmonic hot spot and a molecule with internal vibrational mode (sketched as two masses connected by a spring). Symbols for operators and frequencies are introduced in the text.
(\textbf{B}) During vibrational motion the change in polarizability of the molecule leads to a shift of the plasmon resonance frequency (Eq. \ref{eq:OMshift}) at the origin of the parametric optomechanical coupling.
(\textbf{C}) Schematic molecular potential as a function of the vibrational coordinate. The harmonic oscillator description is valid for small amplitudes (low excitation numbers, dark lines) but anharmonicity must be taken into account under high amplification (higher levels, gray lines).   
}
\end{figure}

Here we show for the first time that SERS scenarios can be  mapped onto the canonical model of cavity optomechanics \cite{Kippenberg2008} (Fig. \ref{fig:Intro}A), in which a {dynamic and coherent interaction takes place between two parametrically coupled and non-resonant harmonic oscillators, namely the molecular vibration and the plasmonic cavity.} The optomechanical coupling rate can be computed from the Raman activity and the plasmonic field distribution, from which we derive the Raman cross section and recover conventional results. 

The novel enhancement mechanism revealed by our approach is dynamical backaction amplification \cite{Braginsky1967} of the vibrational mode due to the non-vanishing plasmon lifetime. In optomechanical microstructures, this effect was first evidenced by the amplification of mechanical breathing modes in silica microtoroids \cite{Kippenberg2005} under blue-detuned excitation, leading to a range of new phenomena as reviewed in \cite{Aspelmeyer_2013}.
We find that SERS systems (i) can feature the suitable dissipation and frequency hierarchies (despite short plasmon decay times, i. e. low quality factors), and (ii) exhibit exceptionally large optomechanical vacuum coupling rates \cite{Gorodetsky}, so that under suitable conditions dynamical backaction amplification can occur via Raman scattering and lead to rich new physics, such as large nonlinearities and out-of-equilibrium vibrational occupancies.

\section*{Model}

\begin{figure}
\begin{centering}
\includegraphics[scale=0.53]{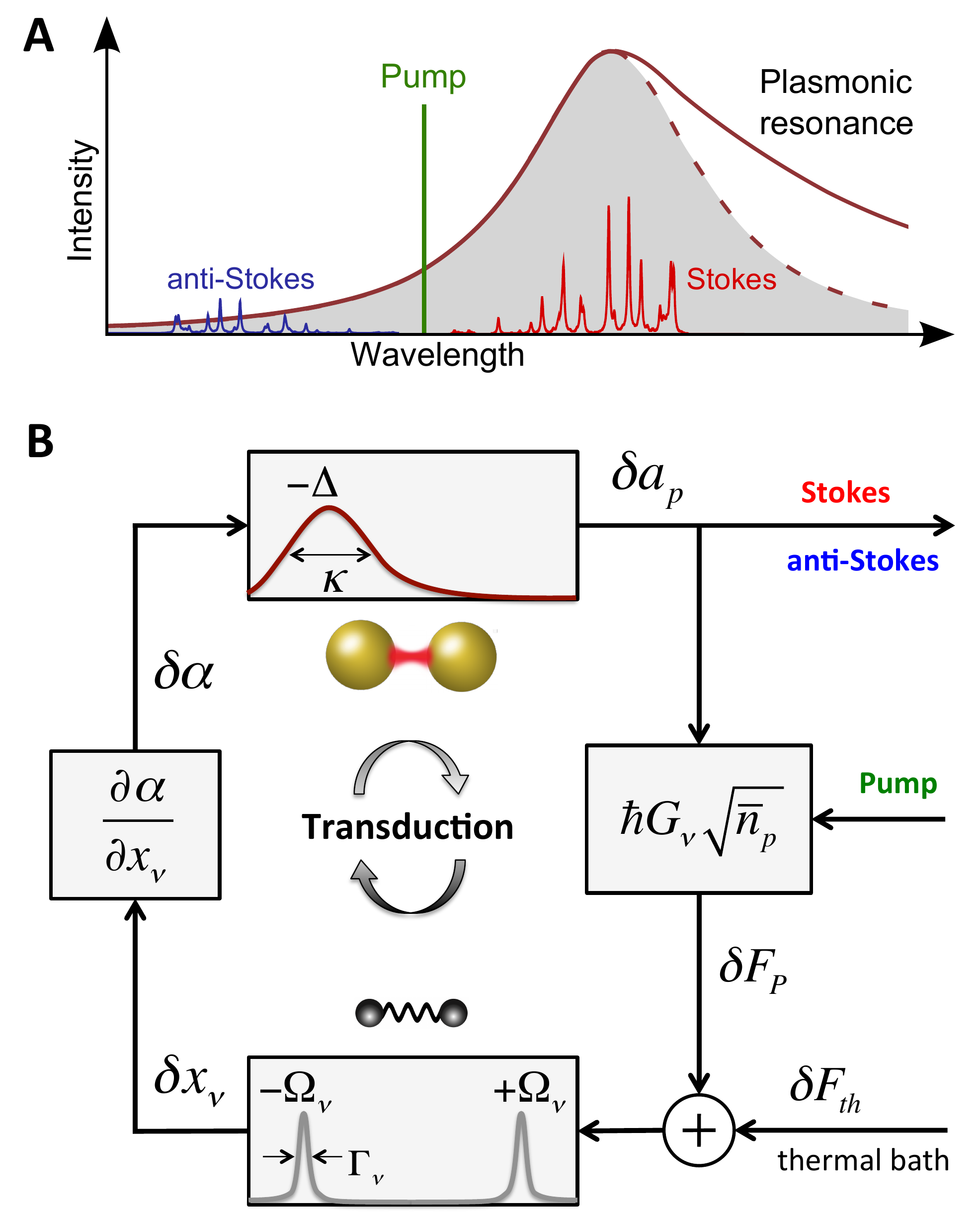}
\par\end{centering}
\protect\caption{\label{fig:Feedback}
\textbf{Feedback diagram of dynamical backaction in the SERS process.}
(\textbf{A}) Schematic Raman spectrum. When the pump is blue-detuned by approximately the vibrational frequency, $\Delta\sim\Omega_{\nu}$, the Stokes process (red) is selectively enhanced over the anti-Stokes process (blue) by the plasmonic resonance (brown line) described by a Lorentzian in the model (shaded area). 
(\textbf{B}) Equivalent feedback diagram of the system. Variables (fluctuations from average) are indicated along the arrows and boxes represent transfer functions. The equations of motion have been linearized and  frequencies are relative to the pump laser (see Supplementary Material).  The pump power controls the amplification factor (via $\bar{n}_P$) in the transduction from the plasmonic field $\delta a_p$ to the force $\delta F_P$ acting on the molecular displacement $\delta x_\nu$. The molecular oscillator acts as a filter at frequencies $\pm \Omega_\nu$.  The displacement is transduced via the Raman activity in a change in polarizability $\delta \alpha$, which modulates the plasmonic field, closing the feedback loop. %The plasmon filter function is centered at $-\Delta$ relative to the laser frequency.
}
\end{figure}

The plasmon mode is formally equivalent to an optical cavity and is modeled by a harmonic oscillator of frequency $\omega_p$ with bosonic creation $\hat{a}^{\dagger}$ and annihilation $\hat{a}$ operators and the Hamiltonian $\hat{H}_p=\hbar\omega_{p}\hat{a}^{\dagger}\hat{a}$ (Fig. \ref{fig:Intro}A). Each normal vibrational mode of a molecule is described by an effective mass $m_{\nu}$, a frequency $\Omega_{\nu} $ and normal coordinates $x_{\nu}$ with the corresponding position operator $\hat{x}_\nu=x_\mathrm{zpm,\nu}(\hat{b}_\nu^{\dagger}+\hat{b}_\nu)$, where $x_\mathrm{zpm,\nu}=\sqrt{\frac{\hbar}{2m_{\nu}\Omega_{\nu}}}$ is the zero-point motion and $\hat{b}_\nu^{\dagger}$, $\hat{b}_\nu$ are bosonic creation and annihilation operators. 
The vibrational mode thermal occupancy at temperature $T$ is given by $\bar{n}_{\nu}=\langle\hat{b}_\nu^{\dagger}\hat{b}_\nu\rangle=(\exp(\hbar\Omega_\nu/k_B T)-1)^{-1}$  ($k_B$ is the Boltzmann constant) which for high frequency Raman-active modes (1000~cm$^{-1}$ corresponding to $\Omega_{\nu}/2\pi=30$~THz) is well below 1 at room temperature (even under moderate heating of the plasmonic particles). The few lowest energy levels of the molecular vibration are well approximated by a harmonic oscillator ($\hat{H}_\nu=\hbar\Omega_{\nu}\hat{b}_{\nu}^{\dagger}\hat{b}_{\nu}$) ( Fig. \ref{fig:Intro}C; possible effects of anharmonicity are discussed later).

Since the frequency hierarchy $\omega_{p}\gg\Omega_{\nu}$ is satisfied {, and assuming the molecule has no optically allowed electronic transitions resonant with the plasmon}, the coupling between vibrational and plasmonic modes is purely \textit{parametric} (as in the case of optomechanical systems \cite{vanLaer2015} and in contrast to resonant coupling \cite{ebbesen,Long_2015}),
and the vibrational displacement leads to a dispersive shift in the plasmon resonance frequency (Fig. \ref{fig:Intro}B) according to 
\begin{equation}\label{eq:OMshift}
\omega_{p}(x_\nu)=\omega_{p}-G_{\nu}\cdot x_{\nu}
\end{equation}
Denoting by $\alpha$ the polarizability of the molecule, which is dependent on the vibrational mode displacement $x_{\nu}$, the optomechanical coupling rate is given by (cf. Supplementary Material)
\begin{equation}
G_{\nu}=\omega_{p}\left(\frac{d\alpha}{dx_{\nu}}\right)\frac{1}{V_{m}\epsilon_{0}}\label{eq:coupling rate}
\end{equation}
with $V_m$ the mode volume of the plasmonic cavity \cite{koenderink} and $\epsilon_{0}$ the permittivity of vacuum. The corresponding vacuum optomechanical coupling rate $g_{\nu,0}=G_{\nu}x_\mathrm{zpm,\nu}$ describes the plasmon frequency shift related to the zero-point motion of the molecular vibration. 

The Hamiltonian for the complete system $\hat{H}=\hat{H}_p+\hat{H}_\nu+\hat{H}_{int}$ is formally identical to the one obtained in cavity optomechanics, with an interaction term \cite{Aspelmeyer_2013}
\begin{equation}\label{eq:Hint}
\hat{H}_{int}=-\hbar\hat{a}^{\dagger}\hat{a}\cdot g_{\nu,0}(\hat{b}_\nu^{\dagger}+\hat{b}_\nu)
\end{equation}
{ that describes the coherent coupling between a mechanical oscillator (here the molecular vibration) and an electromagnetic cavity mode (here the localized plasmon).} The vibration acts on the plasmon via the dispersive plasmon frequency shift of eq. (\ref{eq:OMshift}) that can change the plasmon occupancy $n_P(t)$. In turn, the plasmon acts back on the vibration via the \textit{time-dependent} force
\begin{equation}\label{FRP}
F_{P}(t) =\hbar G_\nu n_P\left(t\right)
\end{equation}
where $n_P\left(t\right)= \langle\hat{a}^{\dagger}(t)\hat{a}(t)\rangle$ is the plasmon occupancy.  The delay in the plasmonic cavity response to changes in the resonance frequency leads to a component of the force that is out of phase with respect to the vibrational motion (see Supplementary Material). When the driving laser frequency matches a phonon sideband ($|\Delta| \sim \Omega_\nu$) the phase shift is $\pm\ \pi/2$, corresponding to a purely viscous force. For blue detuning (Fig. \ref{fig:Feedback}A), this leads to an effective mechanical gain of electromagnetic origin and to parametric amplification. The resulting dynamical backaction can be described as a delayed feedback loop (Fig. \ref{fig:Feedback}B) \cite{Botter_2012}. In the frequency domain, the plasmonic and vibrational time responses correspond to spectral filtering. The transduction from plasmon to vibration occurs through the force $F_P$, with an amplification factor proportional to the intracavity plasmonic field $\sqrt{\bar{n}_P}$ (induced by the pump laser) and to the Raman polarizability $\frac{d\alpha}{dx_{\nu}}$ (contained in $G_\nu$). 

Following the conventions in Raman spectroscopy we introduce the  \textit{mass-weighted cartesian displacement coordinates}\cite{wilson} $Q_{\nu}=\sqrt{m_\mathrm{eff,\nu}}x_{\nu}$.
In this notation the vacuum coupling rate is
\begin{equation}
g_{\nu,0}=\omega_{p}\left(\frac{\partial\alpha}{\partial Q_{\nu}}\right)\left(\frac{1}{V_{m}\epsilon_{0}}\right)\sqrt{\frac{\hbar}{2\Omega_{\nu}}}
\end{equation}
The Raman activity $R_{\nu}$ in a simplified one dimensional model satisfies $R_{\nu}=\left(\frac{\partial\alpha}{\partial Q_{\nu}}\right)^{2}$. Even for small molecules, and without considering possible enhancement from electronic resonances, we obtain optomechanical coupling rates in the 10 -- 100 GHz range (Supplementary Material, section 4), which are  4 to 5 orders of magnitudes higher than in state-of-art microfabricated optomechanical structures \cite{Chan_2012}.  %We also present in the Supplementary Material order-of-magnitude estimates of the coupling rate for the $G$-band of a section of graphene or carbon nanotube, evidencing the promise of these systems \cite{Schedin_2010,Kneipp_2000,Ghamsari_2015} to investigate novel phenomena in molecular cavity optomechanics.

For a realistic value of the plasmon quality factor ($Q\sim10$) \cite{LeRuCh6} in the near-infrared frequency range ($\omega_{p}/2\pi\sim330$~THz for $\lambda=900$~nm)\footnote{Note that for strongly dissipative cavities there is a shift between near- and far-field resonance frequencies ($\omega_{NF}$ and $\omega_{FF}$, respectively). Based on Nordlander’s paper \cite{Nordlander2011}, we can consider this shift as negligible for the type of structures we consider. In addition, numerical studies for a single silver nanosphere and for gold dimers have been carried out elsewhere, showing no sizeable shift between the near and far field spectra \cite{Moreno2013,Ruben2015}.} and a typical linewidth ($\sim2$~cm$^{-1}$ \cite{boyd}) and frequency ($1000$~cm$^{-1}$) of the Raman-active vibrational modes, one obtains a plasmonic dissipation rate $\kappa/2\pi=33$~THz,  a vibrational dissipation rate $\Gamma_{\nu}/2\pi=0.06$~THz and a vibrational frequency $\Omega_{\nu}/2\pi=30$~THz. SERS systems can thereby satisfy the dissipation hierarchy $\Gamma_{\nu}\ll\kappa$. Most importantly, despite the short plasmon lifetime $2\pi/\kappa=30$~fs, there can be sufficient retardation compared to the vibrational period $2\pi/\Omega_\nu=33$~fs for efficient dynamical backaction amplification.%, i.e. $\Omega_\nu \approx \kappa/2$ 

\section*{Raman cross-section calculations}

{ Assuming that the amplitude of vibration is small, the interaction between the plasmonic and the vibrational mode is linearized and} the plasmon-enhanced Raman scattered power is then derived by solving { classical} Langevin equations, which account for both unitary evolution and dissipative terms. When the dissipation hierarchy $\Gamma_\nu \ll \kappa$ is satisfied, and \textit{neglecting backaction} for now (as in conventional SERS theories), the Stokes (resp. anti-Stokes) power is (see Supplementary Material)
\begin{subequations}
\begin{align}
P_{S}/\hbar\omega_L&=\left|\frac{1}{2}x_{\nu}G_{\nu}\left(\frac{\sqrt{\eta\kappa}}{-i\Delta+\kappa/2}\right)\left(\frac{s\sqrt{\eta\kappa}}{(\Omega_{\nu}-\Delta)-i\kappa/2}\right)\right|{}^{2}(\bar{n}_{\nu}+1)\\
P_{AS}/\hbar\omega_L&=\left|\frac{1}{2}x_{\nu}G_{\nu}\left(\frac{\sqrt{\eta\kappa}}{-i\Delta+\kappa/2}\right)\left(\frac{s\sqrt{\eta\kappa}}{(\Omega_{\nu}+\Delta)-i\kappa/2}\right)\right|{}^{2}\bar{n}_{\nu}
\end{align}
\end{subequations}
where the incoming photon flux is
$|s|^{2}=P_\mathrm{in}/\hbar\omega_{L}$ ($P_\mathrm{in}$ being the laser power) and the average occupation number of the mode $\nu$ in equilibrium with an environment at temperature $T$ is $\bar{n}_{\nu}$.  $\eta$ is the fraction of plasmonic energy decaying into  radiation coupled to the excitation/detection optics.
The Stokes (resp. anti-Stokes) Raman cross-section is defined by $\sigma_{S/AS}=P_{S/AS}/I$ where $I=P_\mathrm{in}/A_\mathrm{eff}$ is the laser intensity at the excitation spot of area $A_\mathrm{eff}$.  We obtain the cross-sections for Stokes $\sigma_{S}^\mathrm{em}$ and anti-Stokes $\sigma_{AS}^\mathrm{em}$ scattering \textit{without backaction} (as in conventional electromagnetic theory)
\begin{subequations}
\begin{align}
\sigma_{S}^\mathrm{em}&=\frac{\eta A_\mathrm{eff}}{4}\left(\frac{g_{\nu,0}}{\kappa}\right)^2\left(\frac{1}{\Delta^{2}/\kappa^{2}+1/4}\right)\left(\frac{1}{(\Delta-\Omega_{\nu})^{2}/\kappa^{2}+1/4}\right)(\bar{n}_{\nu}+1)\\
\sigma_{AS}^\mathrm{em}&=\frac{\eta A_\mathrm{eff}}{4}\left(\frac{g_{\nu,0}}{\kappa}\right)^2\left(\frac{1}{\Delta^{2}/\kappa^{2}+1/4}\right)\left(\frac{1}{(\Delta+\Omega_{\nu})^{2}/\kappa^{2}+1/4}\right)\bar{n}_{\nu}
\end{align}\label{eq:sigmat}
\end{subequations}
{To see how this formalism (without backaction yet) recovers the conventional theory, we note that the field inside the plasmonic cavity $E_\mathrm{loc}$ satisfies the energy relation $P_\mathrm{in}\kappa^{-1}\sim\frac{1}{2}\epsilon_{0}|E_\mathrm{loc}|^{2}V_{m}$. On the other hand the incoming power flux is given by $P_\mathrm{in}{A_\mathrm{eff}}^{-1}=\frac{1}{2}\epsilon_{0}c|E_\mathrm{in}|^{2}$, so that we can approximate the field enhancement factor by $\frac{|E_\mathrm{loc}|^{2}}{|E_\mathrm{in}|^{2}} ~\simeq \frac{cA_\mathrm{eff}}{\kappa V_{m}}$}, and
because $g_{\nu,0}\propto 1/V_m$ we find that  $\left(g_{\nu,0}/\kappa\right)^2 \propto\left|\frac{E_\mathrm{loc}}{E_\mathrm{in}}\right|^4$. %represents the local field enhancement factor { in equations (\ref{eq:sigmat}a-b)}. 
This expression recovers the accepted electromagnetic enhancement (``$E^{4}$ law"  \cite{camden,Kneipp1997}) while explicitly showing the contribution of the {plasmonic density of states} at the laser and Stokes (resp. anti-Stokes) frequencies (Fig.~\ref{fig:comparison}A). Because the optomechanical coupling rate is derived from the Raman tensor, our model can be extended to account for all effects previously studied such as (i) the dependence of cross section on the molecule spatial orientation, (ii) the selection rules corresponding to different light polarization, and (iii) the increase in cross section close to electronic resonances, possibly altered by chemical effects and charge transfers.

\section*{A new enhancement mechanism: Dynamical Backaction}

The optomechanical theory uncovers a novel enhancement mechanism: dynamical backaction between the molecular vibration and the plasmon resonance, which leads to a modification of the damping rate { of the vibrational mode}. Under weak coupling condition ($\kappa\gg G_{\nu}$), {the change in damping rate} can be expressed by \cite{Schliesser_2006}
\begin{equation}\label{Gdba}
\Gamma_\mathrm{dba}=4g_{\nu,0}^{2}\bar{n}_{p}\left[\frac{\kappa}{(\Delta+\Omega_{\nu})^{2}+\kappa^{2}/4}-\frac{\kappa}{(\Delta-\Omega_{\nu})^{2}+\kappa^{2}/4}\right]
\end{equation}
where $\bar{n}_{p}=\frac{P_\mathrm{in}}{\hbar\omega_L}\frac{\eta\kappa}{\Delta^{2}+\kappa^{2}/4}$ is the plasmon occupancy.
When pumping the plasmon with blue-detuned light ($\Delta>0$), the backaction damping rate is negative ($\Gamma_\mathrm{dba}<0$) leading to amplification \cite{Kippenberg2005} of the vibrational mode and to an out-of-equilibrium vibrational occupancy
\begin{equation}
n^\mathrm{dba}_{\nu}=\frac{\bar{n}_{\nu}}{1-C}
\end{equation} where $C$ is the cooperativity 
\begin{equation}
C=\frac{\left|\Gamma_\mathrm{dba}\right|}{\Gamma_{\nu}}\label{threshold}
\end{equation}
For  $0<C<1$ the Stokes and anti-Stokes cross-sections are enhanced according to
\begin{subequations}
\begin{align}
\sigma_{S}^\mathrm{dba}&=\frac{n^\mathrm{dba}_{\nu}+1}{\bar{n}_{\nu}+1}\sigma_{S}^\mathrm{em}\\
\sigma_{AS}^\mathrm{dba}&=\frac{n^\mathrm{dba}_{\nu}}{\bar{n}_{\nu}}\sigma_{AS}^\mathrm{em}=\frac{1}{1-C}\sigma_{AS}^\mathrm{em}
\end{align}
\end{subequations}
and they become power dependent, leading to a superlinear increase of the Raman signal with pump intensity%, which is more pronounced as the threshold is approached ($C\rightarrow 1$)
. Also the anti-Stokes/Stokes ratio $R$ becomes ``anomalous" under dynamical backaction (i.e. exhibits values deviating from the equilibrium Boltzmann factor)
\begin{equation}\label{Ratio}
R=\frac{\sigma_{AS}^\mathrm{dba}}{\sigma_{S}^\mathrm{dba}}=\frac{\sigma_{AS}^\mathrm{em}}{\sigma_{S}^\mathrm{em}}\frac{\bar{n}_{\nu}+1}{\bar{n}_{\nu}+1-C} \simeq \frac{\sigma_{AS}^\mathrm{em}}{\sigma_{S}^\mathrm{em}}\frac{1}{1-C}
\end{equation}%>e^{-\frac{\hbar\Omega_\nu}{k_B T}}
where the last approximation is valid for low thermal occupancy typical of high-frequency vibrations. A parametric instability occurs when the amplification rate exceeds the damping rate ($\Gamma_\mathrm{dba}<-\Gamma_{\nu}$),
corresponding to coherent regenerative oscillations (phonon lasing) \cite{Kippenberg2005}. By increasing the quality factor of the plasmonic structure, the threshold is lowered and occurs for a laser detuning closer to the phonon sideband ($\Delta\sim\Omega_\nu$). Close to this threshold the system exhibits a highly nonlinear response, which could provide a mean to achieving super-resolution in TERS \cite{Zhang2013} (see Section 4 \& 5 in the Supplementary Material).

{ Molecular vibrations exhibit anharmonic potentials at higher occupancy (Fig. \ref{fig:Intro}C) and inter-mode couplings that cause internal vibrational redistribution (IVR) \cite{kenkre}, both of which are not included explicitly in our model. IVR would prevent reaching the threshold for regenerative oscillations of a single mode by introducing additional damping channels, and at the same time could lead to the appearance of a broad Raman background originating from other vibrational or rotational modes indirectly excited \cite{Zhang2013}. Anharmonicity of the potential, on the other hand, could lead to frequency shifts and broadening of the Raman peak under high amplification (close to threshold). We note that although the reduced damping rate $\Gamma_{\nu}+\Gamma_\mathrm{dba}$ should manifest as a reduced linewidth of the vibrational mode, this signature could be masked by both IVR and anharmonicity. Observation of the optical spring effect (a static shift in the frequency of the vibration due to the average value of the radiation pressure force) would also be difficult with molecules.}

\subsection*{Collective coupling}

\begin{figure}[h!]
\vspace{-6pt}
\begin{centering}
\includegraphics[scale=0.7]{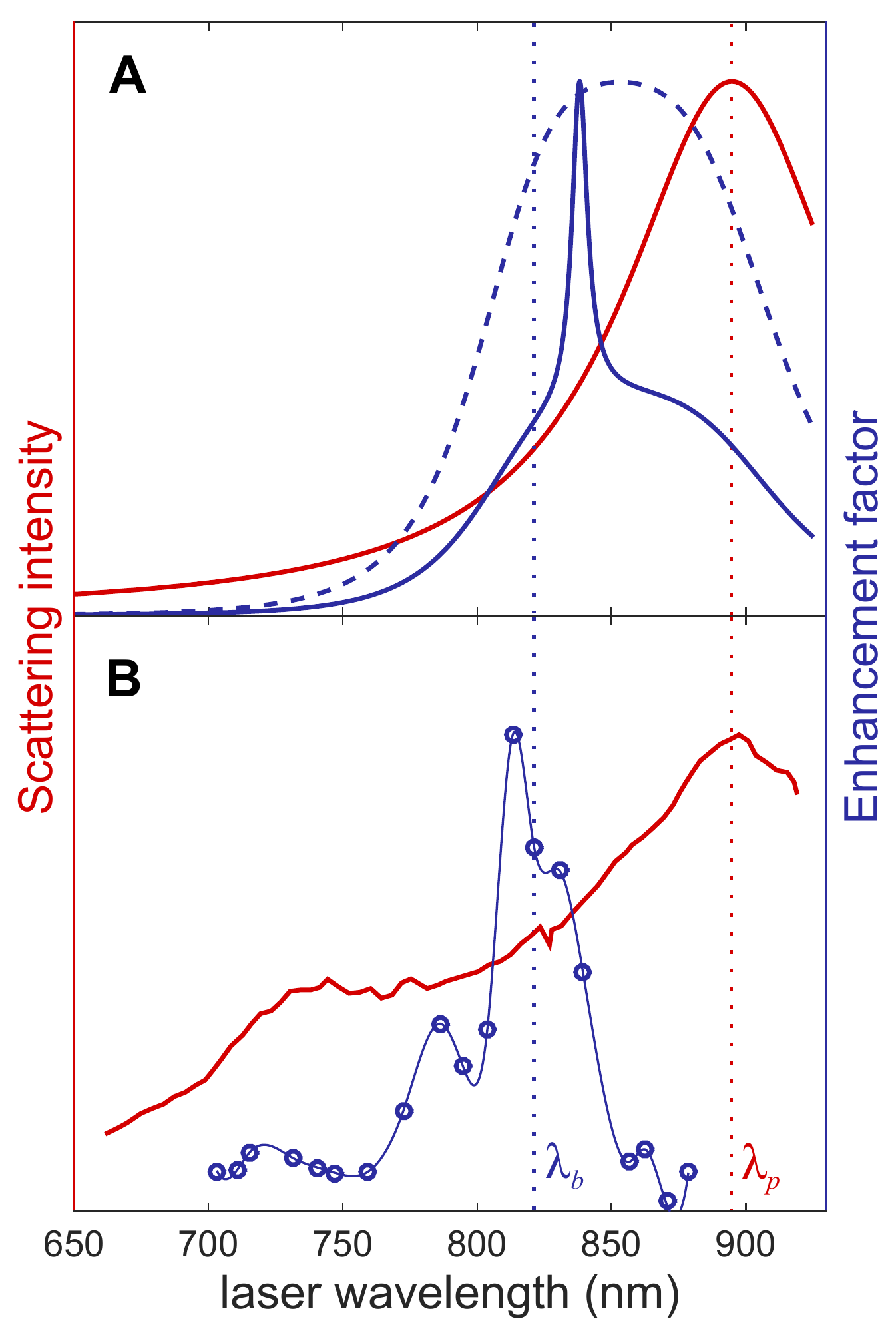}
\par
\end{centering}
\vspace{-6pt}
\protect\caption{\label{fig:comparison}
\textbf{Sharpening of the Raman excitation spectral linewidth}
(\textbf{A}) Calculated Raman enhancement factors as a function of the excitation wavelength without dynamical backaction (conventional theory, blue dashed line) and with dynamical backaction (our prediction, blue solid line) for the parameters used in Ref. \cite{Zhu_2014}, Fig. 3B. The red line shows the plasmon scattering spectrum (arbitrary units) used in our calculations. (\textbf{B}) Experimental results { reprinted with  permission} from Zhu \textit{et al.} \cite{Zhu_2014} for comparison. The open circles are the measured  Raman enhancement factors while the red line show the experimental plasmon scattering spectrum. {The red dotted vertical line indicates the position of the plasmon resonance ($\lambda_p$) and the blue dotted vertical line the position of the blue (anti-Stokes) phonon sideband ($\lambda_b$).}
}
\end{figure}

The innovative fabrication process and careful characterization methods described by Zhu and Crozier in \cite{Zhu_2014} enable the authors to investigate plasmonic cavities with well-defined field confinement properties, reaching down to the onset of quantum tunneling between the two metallic parts \cite{Savage,scholl2013observation}. Based on the geometric estimate of the field confinement described in \cite{Savage}, we can approximate the mode volume in \cite{Zhu_2014} by $V_m\simeq2.0\cdot10^{-7}\lambda^{3}$ (see Supplementary Material, Section 3.2). This volume also sets the approximate number of molecules $N$ contributing to the plasmon-enhanced Raman signal. Considering monolayers of thiophenol covering the metallic cylinders, we find $N\sim \rho_S\ r^{1/2}d^{1/2}h$ with $\rho_S$ the thiophenol surface density, $r$ the radius of the cylinder, $h$ its height and $d$ the size of the gap. 
Under these conditions the optomechanical interaction should be coherently driving a collective oscillator mode \cite{kipf}.\footnote{Because the plasmon lifetime in the cavity ($<0.1$ ps) is much faster than the pure dephasing of molecular vibrations ($>10$ ps, as estimated from the Raman linewidth) we believe that dephasing would not prevent coherent driving of the collective vibrational mode.} In the Supplementary Material, section 1.5, we show that the optomechanical coupling rate between this self-assembled monolayer and the plasmon is enhanced as: $g_{\nu,0}=\sqrt{N}G_{\nu}x_\mathrm{zpm,\nu}$.

Introducing the experimental parameters (at the onset of quantum tunneling) and assuming $\eta\sim 3/4$ \cite{giessen} in our calculations, we estimate that Zhu and Crozier are reaching the parametric amplification instability $C=1$ for an incoming power of $10\ \mu$W. This compares favorably to the value of $20\ \mu$W below which the authors operate in order to avoid sample damages and non-linear effects in the plasmon. Consequently, we predict that their experiment might be showing signatures of dynamical backaction amplification. It is therefore worth comparing our predictions with the conventional theory of SERS. 

Setting the plasmon resonance at a wavelength of $895$~nm and the incoming power slightly below the threshold, we provide a comparison between the two models (without and with dynamical backaction) and the experimental results of Zhu \textit{et al.} (Fig.~\ref{fig:comparison}). Under the condition of negligible dynamical backaction (red dashed line) maximal Raman signal occurs when the pump wavelength is half-way between the plasmon resonance and the phonon sideband, recovering the results of the electromagnetic theory \cite{McFarland_2005}. The sharpening and shift of the Raman excitation spectrum can be qualititavely predicted and explained by our new model that allows dynamical backaction to take place in the plasmonic cavity. %{\color{red} Comment for mismatch between predicted and observed Raman excitation peak?}

\begin{figure}[h!]
%\vspace{-30pt}
\begin{centering}
\includegraphics[scale=0.8]{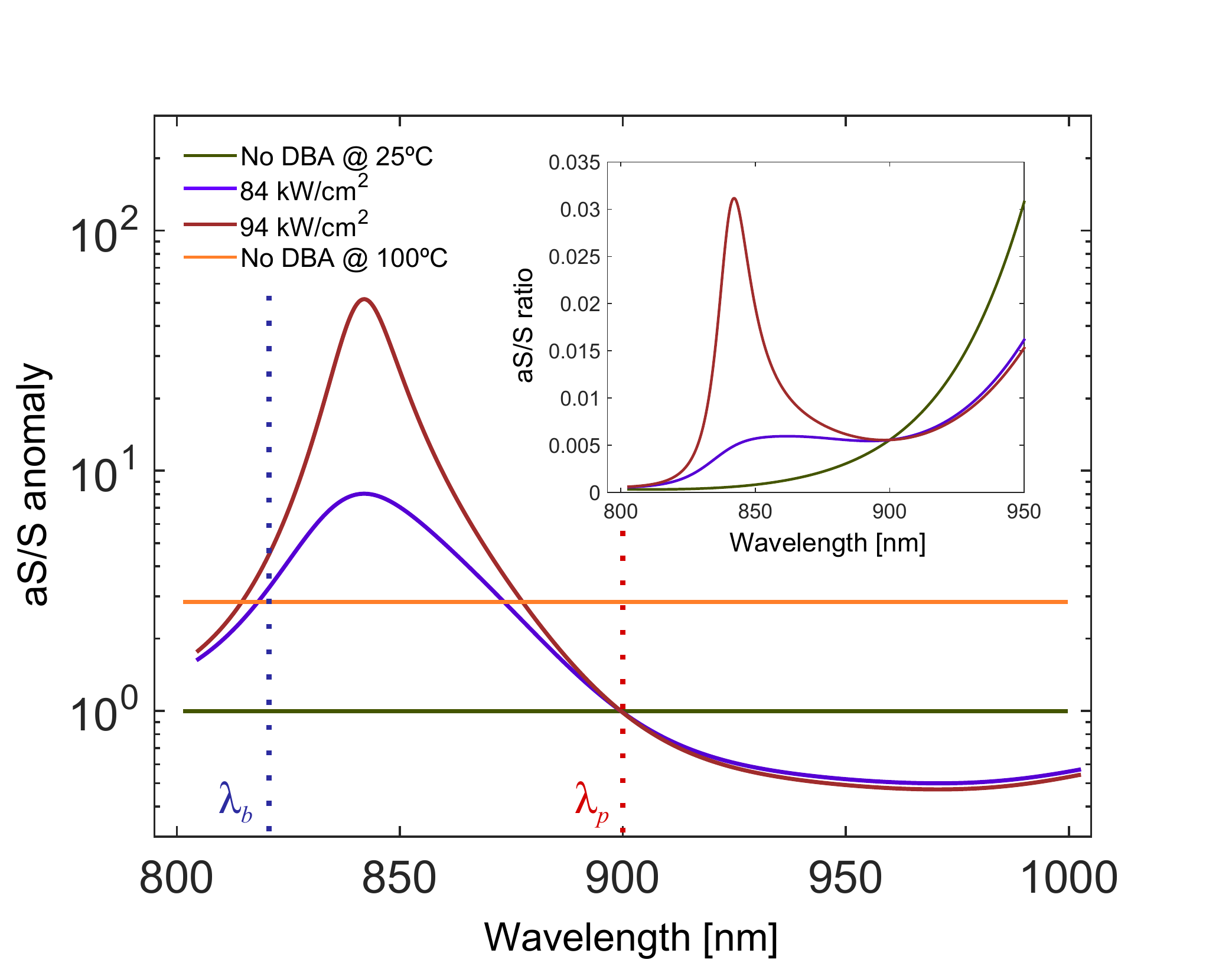}
\par
\end{centering}
%\vspace{-20pt}
\protect\caption{\label{fig:DBA}
\textbf{Anomalous anti-Stokes/Stokes ratio under dynamical backaction amplification.}
The anti-Stokes/Stokes (aS/S) anomaly is defined as the aS/S ratio $R={\sigma_{AS}^\mathrm{dba}}/{\sigma_{S}^\mathrm{dba}}$ (eq. \ref{Ratio}) divided by its value under negligible dynamical backaction amplification (${\sigma_{AS}^\mathrm{em}}/{\sigma_{S}^\mathrm{em}}$) for a vibrational mode in thermal equilibrium at room-temperature (No DBA @ 25$^{\text{o}}$C). Parameters are as in Fig.~\ref{fig:comparison}. { The plasmonic resonance is at $\lambda_p$ and the blue phonon sideband at $\lambda_b$.}
 For comparison we also plot the aS/S ratio when the molecules are heated by the laser to $100^{\text{o}}$C (blue line) { assuming that absorption is wavelength independent. The insert shows the original aS/S ratio $R$ before normalization, as expressed in (eq. \ref{Ratio}), for the three pump intensities considered in the main figure. Even without backaction, the ratio depends on the laser detuning from the plasmon due to selective enhancement of either Stokes or anti-Stokes outgoing wavelengths.}
}
\end{figure}

The anomalous anti-Stokes/Stokes ratio contemplated in the previous section could also be explored with the experimental setup here described. The deviation from the equilibrium Boltzmann factor is demonstrated in Fig. \ref{fig:DBA} for realistic pump intensity. { The signature of dynamical backaction amplification can be experimentally distinguished from thermal effects such as local heating by (i) the sharp sensitivity of the anomaly on the laser detuning from the plasmon; and (ii) its non-linear power dependence (see Fig. S1B in the Supplementary Material). }

%To summarize, the new dynamical backaction mechanism uncovered by our theory of SERS and TERS predicts mode-selective vibrational amplification under blue-detuned excitation, leading in particular to non-equilibrium anti-Stokes/Stokes scattering ratios and to a highly nonlinear additional enhancement of the Raman cross-section close to the instability threshold.

\section*{Outlook}

The theory provides a unified, physically grounded and quantitative framework for the phenomena of SERS.
It challenges the common wisdom, based on the ``$E^4$ law", that broader plasmonic resonances -- overlapping with both laser and Stokes wavelengths -- achieve the highest possible enhancement. Instead we show that vibrational amplification is more efficient for larger plasmon $Q$-factors (i.e. narrower resonances) and for blue detuned excitation on the anti-Stokes vibrational sideband. {We note that even higher amplification factors could be obtained for double plasmonic resonances spaced by the vibrational frequency, corresponding to a multi-mode optomechanical system}. These insights are of major relevance for the design of SERS systems leveraging backaction for more sensitive spectroscopy and higher resolution imaging.

More broadly, the theory lays the foundations of molecular cavity optomechanics and opens unforeseen research directions. The rich physics of cavity optomechanics is now accessible in systems of nanometric dimensions featuring coupling rates several orders of magnitude higher than state-of-the-art microfabricated devices. One example of observable phenomenon is the counter-intuitive reduction of the Stokes cross-section under simultaneous pumping with a second, red-detuned laser, which would lead to damping of the vibrational mode. Other effects include optomechanically-induced transparency and frequency conversion. {Functional materials could be designed using layers of molecules collectively coupled to two-dimensional cavities. By choosing molecules exhibiting vibrational modes that are both infrared and Raman active, one could achieve coherent frequency conversion between THz radiation and visible light.} Finally, because of the high frequency of molecular vibrational modes, they are not thermally excited at room temperature, fostering the prospects for molecular \textit{quantum} optomechanics. With the mechanical oscillator in its quantum ground state, schemes such as the creation of non-classical states of motion \cite{Galland_2014} become feasible with simple experimental setups at room temperature.

\textbf{Acknowledgments}
The authors would like to thank Dr. E. Verhagen for stimulating discussions
and Dr. E. Bremond for his precious help in running the chemical simulations.
This work was partially supported via an ERC Advanced Grant, NCCR
of Quantum Engineering, QSIT as well as the Swiss National Science Foundation. C.~G. acknowledges the support of the Swiss National Science Foundation through an \textit{Ambizione} Fellowship. N.~P. acknowledges the support of a Marie-Curie Fellowship.

\pagebreak

%%%%%%%%%% Merge with supplemental materials %%%%%%%%%%
\widetext
\clearpage
\begin{center}
\textbf{\large -- Supplementary Material -- \\Molecular cavity optomechanics: a theory of plasmon-enhanced Raman scattering}
\end{center}
%%%%%%%%%% Merge with supplemental materials %%%%%%%%%%
%%%%%%%%%% Prefix a "S" to all equations, figures, tables and reset the counter %%%%%%%%%%
\setcounter{equation}{0}
\setcounter{figure}{0}
\setcounter{table}{0}
\makeatletter
\renewcommand{\theequation}{S\arabic{equation}}
\renewcommand{\thefigure}{S\arabic{figure}}
\renewcommand{\thetable}{S\Roman{table}}
\renewcommand{\bibnumfmt}[1]{[S#1]}
\renewcommand{\citenumfont}[1]{S#1}
%%%%%%%%%% Prefix a "S" to all equations, figures, tables and reset the counter %%%%%%%%%%

%\begin{refsection}

\section{Cavity optomechanical principles}

\subsection{Symbols and definitions}
\begin{center}
\begin{table}[ht!]
\begin{centering}
\begin{tabular}{|c|c|}
\hline 
Notation & Description\tabularnewline
\hline 
\hline 
$a$ & Dimensionless intracavity field (normalized so that $|a|^2=$photon number) \tabularnewline
\hline 
$s$ & Flux of the pump field in units of s$^{-1/2}$ ($|s|^2$ is the photon flux)\tabularnewline
\hline 
$\omega_L/2\pi$ & Frequency of the excitation laser \tabularnewline
\hline 
$\omega_p/2\pi$ & Resonance frequency of the plasmonic cavity\tabularnewline
\hline 
$\kappa/2\pi$ & Total energy damping rate of the plasmonic cavity\tabularnewline
\hline 
$\kappa_0/2\pi$ & Rate of intrinsic losses, including absorption and uncoupled radiation\tabularnewline
\hline 
$\kappa_{ex}/2\pi$ & Coupling rate between the cavity field and the incoming/outgoing fields \tabularnewline
\hline 
$\kappa=\kappa_{0}+\kappa_{ex}$ & Total decay rate of the plasmonic cavity\tabularnewline
\hline 
$\eta=\kappa_{ex}/\kappa$ & Coupling ratio ($\eta=1/2$ for critical coupling)\tabularnewline
\hline
$\Gamma_{\nu}/2\pi$ & Energy damping rate of the vibrational mode $\nu$\tabularnewline
\hline 
$\Omega_{\nu}/2\pi$ & Mechanical frequency of the vibrational mode $\nu$\tabularnewline
\hline 
$x_{\nu}$ & Displacement coordinate of the vibrational mode $\nu$\tabularnewline
\hline 
$m_{\nu}$ & Effective mass of the vibrational mode $\nu$\tabularnewline
\hline 
$x_\mathrm{zpm,\nu}$ & Zero-point motion of the vibrational mode $\nu$\tabularnewline
\hline 
$\bar{n}_{\nu}$ & Average occupation number of the vibrational mode $\nu$\tabularnewline
\hline 
$\bar{n}_{p}$ & Average number of plasmons inside the cavity\tabularnewline
\hline 
$\Delta=\omega_L-\omega_p$ & Detuning between the laser and the plasmon
resonance frequency\tabularnewline
\hline 
$G_{\nu}/2\pi$ & Coupling rate between the plasmon and the vibrational mode $\nu$\tabularnewline
\hline 
$g_{\nu,0}/2\pi$ & Optomechanical vacuum coupling rate to the vibrational mode $\nu$\tabularnewline
\hline 
\end{tabular}
\par\end{centering}

\end{table}

\par\end{center}

\subsection{From the formal Hamiltonian to classical equations of motion}

We present in the following a simple treatment of the optomechanical system to give an intuitive picture of the important phenomena considered in the manuscript. A complete derivation including all the assumptions made can be found in reviews on cavity optomechanics \cite{Aspelmeyer_2013,Kippenberg_07}. \\

We start from the expression describing the interaction between a
radiation mode and a vibrational mode $\nu$ as introduced in the main text
\begin{equation}
\hat{H_{0}}=\hbar\omega_{p}\hat{a}^{\dagger}\hat{a}+\hbar\Omega_{\nu}\hat{b}^{\dagger}\hat{b}-\hbar G_{\nu}x_\mathrm{zpm,\nu}\hat{a}^{\dagger}\hat{a}\left(\hat{b}^{\dagger}+\hat{b}\right)
\end{equation}
(We recall the expression of the zero-point motion: $x_\mathrm{zpm,\nu}= \sqrt{\frac{\hbar}{2m_{\nu}\Omega_{\nu}}}$). 

A first derivation of this optomechanical Hamiltonian was given by
Law \cite{law}. The interaction term is corresponding to a nonlinear
process that involves a product of three operators. The driving laser at frequency $\omega_{L}$ is modeled by a coherent field:
$\hat{H}_{L}=i\hbar s\sqrt{\kappa_{ex}}\left(e^{-i\omega_{L}t}\hat{a}^{\dagger}-e^{i\omega_{L}t}\hat{a}\right)$. It is convenient to switch to a reference frame rotating at
the laser frequency, in which the total Hamiltonian writes
\begin{equation}
\hat{H}=e^{i\omega_{L}\hat{a}^{\dagger}\hat{a}t}\left(\hat{H_{0}}+\hat{H}_{L}\right)e^{-i\omega_{L}\hat{a}^{\dagger}\hat{a}t}-\hbar\omega_{L}\hat{a}^{\dagger}\hat{a}
\end{equation}
{Making use of the Baker-Campbell-Hausdorff formula, this change of reference frame allows to eliminate the explicit time dependence of the Hamiltonian.}
Dissipations can be introduced by writing the quantum Langevin equations. Within the input-output formalism \cite{gardiner}, defining $\Delta=\omega_L-\omega_p$ as the detuning between the laser and the plasmon, we obtain the equations
\vspace{6pt}
\begin{eqnarray}\label{QLE}
\dot{\hat{a}} & = & -\frac{i}{\hbar}\left[\hat{a,}\hat{H}\right]-\frac{\kappa_{0}}{2}\hat{a}+\sqrt{\kappa_{0}}\hat{a}_{in,0}-\frac{\kappa_{ex}}{2}\hat{a}+\sqrt{\kappa_{ex}}\hat{a}_{in,ex}\nonumber \\
 & = & i\Delta\hat{a}+iG_{\nu}\hat{a}\left(\hat{b}^{\dagger}+\hat{b}\right)x_\mathrm{zpm,\nu}-\frac{\kappa}{2}\hat{a}+\sqrt{\kappa_{0}}\hat{a}_{in,0}+\sqrt{\kappa_{ex}}\left(s+\hat{a}_{in,ex}\right)\\
\dot{\hat{b}} & = & -\frac{i}{\hbar}\left[\hat{b,}\hat{H}\right]-\frac{\Gamma_{\nu}}{2}\hat{b}+\sqrt{\Gamma_{\nu}}\hat{b}_{in}\nonumber \\
 & = & -i\Omega_{\nu}\hat{b}+iG_{\nu}\hat{a}^{\dagger}\hat{a}x_\mathrm{zpm,\nu}-\frac{\Gamma_{\nu}}{2}\hat{b}+\sqrt{\Gamma_{\nu}}\hat{b}_{in}
\end{eqnarray}
where we introduced the input noise terms expressed in the rotating frame: the vacuum noise $\hat{a}_{in,0}$ and driving laser's noise $\hat{a}_{ex,0}$ entering the plasmonic cavity, and the thermal noise $\hat{b}_{in}$ on the vibrational mode. Assuming a shot-noise limited laser and neglecting the thermal excitation at the plasmon frequency, the correlators associated with these fluctuations are given by:
\begin{eqnarray}
\left\langle \hat{a}_{in}^{\dagger}(t)\hat{a}_{in}(t')\right\rangle & = & 0\\
\left\langle \hat{a}_{in}(t)\hat{a}_{in}^{\dagger}(t')\right\rangle & = & \delta(t-t')\\
\left\langle \hat{b}^{\dagger}_{in}(t)\hat{b}_{in}(t')\right\rangle & = & \bar{n}_{\nu}\delta(t-t')\\
\left\langle \hat{b}_{in}(t)\hat{b}^{\dagger}_{in}(t')\right\rangle & = & (\bar{n}_{\nu}+1)\delta(t-t')
\end{eqnarray}
where $\bar{n}_{\nu}$ is the thermal occupancy of the bath at the vibrational frequency $\Omega_{\nu}$ (valid for sufficiently high vibrational quality factor).

The equations of motion for the corresponding creation operators are obtained from the
relations $\dfrac{\mathrm{d}}{\mathrm{d}t}\left(\hat{a}^{\dagger}\right)=\left(\dfrac{\mathrm{d}\hat{a}}{\mathrm{d}t}\right)^{\dagger}$,
 $\dfrac{\mathrm{d}}{\mathrm{d}t}\left(\hat{b}^{\dagger}\right)=\left(\dfrac{\mathrm{d}\hat{b}}{\mathrm{d}t}\right)^{\dagger}$. These
equations are describing the complete evolution of the plasmonic excitations coupled
to the molecular vibrations. It is  instructive
to give first the classical version of these equations. \\

We express the vibrational degree-of-freedom in the position and momentum
operators defined by
\vspace{-12pt}
\begin{eqnarray}
\hat{x}_{\nu} & = & x_\mathrm{zpm,\nu}\left(\hat{b}^{\dagger}+\hat{b}\right)\\
\hat{p}_{\nu} & = & ix_\mathrm{zpm,\nu}m_{\nu}\Omega_{\nu}\left(\hat{b}^{\dagger}-\hat{b}\right)
\end{eqnarray}
In this representation the Langevin equations  for the vibrational mode become
\begin{eqnarray}
\dfrac{\mathrm{d}\hat{x}_{\nu}}{\mathrm{d}t} & = & \frac{\hat{p}_{\nu}}{m_{\nu}}-\frac{\Gamma_{\nu}}{2}\hat{x}_{\nu}+\sqrt{\Gamma_{\nu}}\hat{x}_{in}\\
\dfrac{\mathrm{d}\hat{p}_{\nu}}{\mathrm{d}t} & = & -m_{\nu}\Omega_{\nu}^{2}\hat{x}_{\nu}+\hbar G_{\nu}\hat{a}^{\dagger}\hat{a}-\frac{\Gamma_{\nu}}{2}\hat{p}_{\nu}+\sqrt{\Gamma_{\nu}}\hat{p}_{in}
\end{eqnarray}
where $\hat{x}_{in}$ and $\hat{p}_{in}$ are the quantum noise operators.

%\pagebreak

%{The validity of this assumption requires the optomechanical coupling rate to be small compare to $\kappa$, $\Omega_{\nu}$ and the incoming photon flux $\frac{P_\mathrm{in}}{\hbar\omega_p}$ \cite{Aspelmeyer_2013}. It should be observed that the system considered in this study is well fulfilling those different requirements.}
After taking expectation values the noise terms average to zero and one obtains

\pagebreak

\vspace{6pt}
\begin{eqnarray}
\dot{a}-i(\Delta+G_{\nu}x_{\nu})a+\frac{1}{2}\kappa a & = & \sqrt{\kappa_{ex}}s\label{eq:LEo}\\
\ddot{x}_{\nu}+\Gamma{}_{\nu}\dot{x}_{\nu}+\Omega_{\nu}^{2}x_{\nu} & = & \frac{F_{P}(t)}{m_{\nu}}\label{eq:LEm}
\vspace{6pt}
\end{eqnarray}
where $a=\left\langle \hat{a}\right\rangle $ and $x_{\nu}=\left\langle \hat{x}_{\nu}\right\rangle $.
This force of electromagnetic origin \footnote{This force, sometime called radiation pressure force, is more generally defined as the derivative of the interaction part of the Hamiltonian with respect to displacement: $F_{P}(t)=- \frac{d\hat{H}_{int}}{dx_\nu}$ \vspace{12pt}}
 is $F_{P}=\hbar G_{\nu}|a(t)|^{2}$ with $|a|^{2}=\left\langle \hat{a}^{\dagger}\hat{a}\right\rangle$.

\subsection{Linearization of the cavity field}

Setting the time derivatives to zero in the system of equations described
above, we find the stationary solutions
\begin{equation}
\bar{a}=\frac{\sqrt{\kappa_{ex}}s}{-i\left(\Delta+G_{\nu}\bar{x}_{\nu}\right)+\frac{\kappa}{2}}\qquad\bar{x}_{\nu}=\frac{\hbar G_{\nu}}{m_{\nu}\Omega_{\nu}^{2}}\left|a\right|^{2}
\end{equation}

We assume that the amplitude of the vibrations is described
by $x_{\nu}\left(t\right)=x_{\nu,0}\cos\left(\Omega_{\nu}t\right)$
and is small in comparison to the typical length scale $L$ of the plasmonic
cavity\footnote{A large vibrational motion - arising above the parametric threshold described in the manuscript - would break our linear treatment and the contributions of the non-linear terms should be considered \cite{Marquardt_2006}}. We introduce $\epsilon=\frac{x_{\nu,0}}{L}\ll 1$ and use this small dimensionless parameter to develop the cavity field variable in a power series $a=\sum_{n=0}^{\infty}\epsilon{}^{n}a_{n}$. The power expansion of the equation (\ref{eq:LEo}) yields 
\begin{equation}
\sum_{n=0}^{\infty}\epsilon^{n}\frac{da_n}{dt}=\sum_{n=0}^{\infty}\left[\left(i\Delta-\frac{\kappa}{2}\right)\epsilon^{n}a_{n}+iG_{\nu}L\epsilon^{n+1}a_{n}\cos\left(\Omega_{\nu}t\right)\right]+\sqrt{\kappa_{ex}}s
\end{equation}

\paragraph{$\mathbf{0}^{th}$ order equation}
\begin{equation}
\frac{da_{0}}{dt}=\left(i\Delta-\frac{\kappa}{2}\right)a_{0}+\sqrt{\kappa_{ex}}s
\end{equation}
The homogeneous solution can be neglected if the measurement time is much longer than
the other timescales in the system (as is always the case in SERS). We can thus consider only the
particular solution, equal to the steady-state solution without optomechanical coupling
\begin{equation}
a_{0}=\frac{\sqrt{\kappa_{ex}}s}{-i\Delta+\frac{\kappa}{2}}
\end{equation}

\paragraph{$\mathbf{1}^{st}$ order equation}
\begin{equation}
\frac{da_{1}}{dt}=\left(i\Delta-\frac{\kappa}{2}\right)a_{1}+iG_{\nu}La_{0}\cos\left(\Omega_{\nu}t\right)
\end{equation}
We can neglect the homogeneous solution following the same reasoning
as before. We look for a particular solution of the form $a_{1}(t)=a_{AS}e^{i\Omega_{\nu}t}+a_{S}e^{-i\Omega_{\nu}t}$ and obtain the
amplitudes of the anti-Stokes and the Stokes fields, respectively
\begin{equation}\label{Stokesfield}
a_{AS}=\frac{G_{\nu}L}{2}\frac{a_{0}}{\Omega_{\nu}-\Delta-i\frac{\kappa}{2}}\qquad a_S=-\frac{G_{\nu}L}{2}\frac{a_{0}}{\Omega_{\nu}+\Delta+i\frac{\kappa}{2}}
\end{equation}

%\vspace{-12pt}

\subsection{Dynamical Backaction Force}\label{RPforce}

We develop the expression of the force to first order in $\epsilon$
\begin{equation}
F_{P}\left(t\right)=\hbar G_{\nu}|a(t)|^{2}\simeq\hbar G_{\nu}|a_{0}+\epsilon a_{1}(t)|^{2}=\hbar G_{\nu}|a_{0}|^{2}+2\epsilon\hbar G_{\nu}\Re\left(a_{0}a_{1}^{*}(t)\right)
\end{equation}
In the last expression, the first term corresponds to a constant
force applied to the vibrational mode 
$\bar{F}_P=\hbar G_{\nu}|a_{0}|^{2}$. The second term is time-dependent
and can be expressed as a sum of  in- and out-of-phase components
\begin{equation}
\delta F_{P}(t)=\cos\left(\Omega_{\nu}t\right)\delta F_{I}+\sin\left(\Omega_{\nu}t\right)\delta F_{Q}
\end{equation}
with
%\vspace{6pt}
\begin{eqnarray}
\delta F_{I} & = & \hbar G_{\nu}^{2}x_{\nu}|a_{0}|^{2}\left[\frac{\left(\Omega_{\nu}-\Delta\right)}{\left(\Omega_{\nu}-\Delta\right)^{2}+\left(\frac{\kappa}{2}\right)^{2}}-\frac{\left(\Omega_{\nu}+\Delta\right)}{\left(\Omega_{\nu}+\Delta\right)^{2}+\left(\frac{\kappa}{2}\right)^{2}}\right]\\
\delta F_{Q} & = & \hbar G_{\nu}^{2}x_{\nu}|a_{0}|^{2}\left[\frac{\frac{\kappa}{2}}{\left(\Omega_{\nu}+\Delta\right)^{2}+\left(\frac{\kappa}{2}\right)^{2}}-\frac{\frac{\kappa}{2}}{\left(\Omega_{\nu}-\Delta\right)^{2}+\left(\frac{\kappa}{2}\right)^{2}}\right]
%\vspace{6pt}
\end{eqnarray}
\vspace{6pt}

\noindent We can now insert the expression for the electromagnetic force
into the Langevin equation (\ref{eq:LEm})
%\vspace{6pt}
\begin{eqnarray}
m_{\nu}\ddot{x}_{\nu} & = & m_{\nu}\Omega_{\nu}\Gamma_{\nu}x_{\nu,0}\sin\left(\Omega_{\nu}t\right)-m_{\nu}\Omega_{\nu}^{2}x_{\nu,0}\cos\left(\Omega_{\nu}t\right)\nonumber \\
 &  & +\bar{F}_P+\cos\left(\Omega_{\nu}t\right)\delta F_{I}+\sin\left(\Omega_{\nu}t\right)\delta F_{Q}\\
m_{\nu}\ddot{x}_{\nu} & = & -m_{\nu}\Gamma_{\nu}^{'}\dot{x}_{\nu}-m_{\nu}\Omega_{\nu}^{'2}x_{\nu}+\bar{F}_P
%\vspace{6pt}
\end{eqnarray}
%\vspace{6pt}

\noindent The equation is thus describing a mechanical oscillator displaced by a constant force $\bar{F}_P$, with a shifted natural frequency $\Omega_{\nu}^{'}$ and damped at an effective rate $\Gamma_{\nu}^{'}$
  \pagebreak
 %\vspace{6pt}
\begin{eqnarray}
\Gamma_{\nu}^{'} & = & \Gamma_{\nu}+\frac{\delta F_{Q}}{m_{\nu}\Omega_{\nu}x_{\nu,0}}:=\Gamma_{\nu}+\Gamma_\mathrm{dba}\\
\Omega_{\nu}^{'} & = & \sqrt{\Omega_{\nu}^{2}-\frac{\delta F_{I}}{m_{\nu}x_{\nu,0}}}
%\vspace{6pt}
\end{eqnarray}
\vspace{6pt}

\noindent Depending on the laser detuning from the (shifted) plasmonic resonance, three different situations can arise: 
\begin{itemize}
\item $\Delta=0$: For resonant excitation there is no radiation-induced change of the damping rate 
\item $\Delta<0$: For a red-detuned laser, the damping rate is increased, corresponding to ``cooling" 
\item $\Delta>0$: For a blue-detuned laser, the damping rate is decreased, corresponding to  amplification
\end{itemize}
Deep in the sideband-resolved regime ($\kappa\ll\Omega_{\nu}$) the maximal increase (decrease) of the damping rate is obtained when $\Delta=-\Omega_{\nu}$ ($\Delta=\Omega_{\nu}$) and the expression for the maximum backaction damping rate due to optomechanical interactions can be approximated by
\begin{equation}
\Gamma_\mathrm{dba}=\Gamma_{\nu}^{'}-\Gamma_{\nu}=\frac{4\bar{n}_{p}g_{\nu,0}^{2}}{\kappa}
\end{equation}
where $g_{\nu,0}= G_{\nu}\sqrt{\frac{\hbar}{2m_{\nu}\Omega_{\nu}}}$.
When the damping is increased trough the coupling with the plasmonic
field (red detuning), the vibrational mode is losing power into the plasmonic cavity
and thus is being optically ``cooled". On the contrary, when the damping rate is decreased (blue detuning), power is transfered from the plasmonic field to the vibrational mode, whose motion is thereby amplified: this is dynamical backaction amplification.

\subsection{Collective optomechanical plasmon-vibrational coupling}
In this section, we consider $N$ identical phonon modes -- with annihilation operators $\hat{b}_{i=1,...,N}$ -- coupled to the same plasmonic cavity. Corresponding experimental situations include (i) a layer of molecules filling the gap of a metal dimer \cite{Zhu_2014}, where each $\hat{b}_{i}$ represents the same Raman mode for each molecule; (ii) the optical phonon of a piece of bulk or 2D material interacting with a plasmonic cavity, where each $\hat{b}_{i}$ represent the same vibrational mode of each unit cell (for example the $G$-band of graphene in \cite{Ghamsari_2015}). For simplicity, we consider identical vibrational frequencies $\Omega^{\left(i\right)}_{\nu}=\Omega$, intrinsic damping rates $\Gamma^{\left(i\right)}_{\nu}=\Gamma$ and coupling rates to the cavity $G^{\left(i\right)}_{\nu}=G$. 
We follow a derivation similar to \cite{agarwal_2013},  starting from the  Langevin equations in the rotating frame of the molecular vibration and of the Stokes field. Because the plasmon decay rate is much larger than the vibrational damping $\left(\kappa\gg\Gamma\right)$ the cavity field evolution can be adiabatically eliminated and replaced by its steady state solution. This leads to a linear system of $N$ differential equations describing the  evolution of the $N$ phonon modes coupled via the cavity driven by a laser tuned on the anti-Stokes (blue) vibrational sideband:
\begin{equation}
\dot{\hat{b}}_i=-\left(\Gamma+\Gamma_\mathrm{dba}\right)\hat{b}_i+\sum_{j\neq i}\Gamma_\mathrm{dba}\hat{b}_j
\end{equation}
The $N$ solutions of these equations reveal two different behaviours. On the one hand, we find $N-1$ linearly independent and degenerate eigenmodes $\hat{D}_k$ ($k \in \left[1,N-1\right]$) with the same eigenvalue $\left\{-\Gamma\right\}$ that can be written in the general form: 
\begin{equation}
\frac{1}{K_k}\sum_{j}\lambda_{k,j}\hat{b}_j \quad \mathrm{with} \quad K_k=\sqrt{\sum_j \left|\lambda_{k,j}\right|^2}
\end{equation} 
such that $\sum_j \lambda_{k,j}=0 \quad \forall k \in \left[1,N-1\right]$. The shift of the plasmonic resonance caused by each of the collective modes $\hat{D}_k$ is proportional to its collective optomechanical coupling rate $G_k=\sum_j \lambda_j G = 0 $. This shows that these modes are dark, i.e. decoupled from the plasmonic cavity and thus not affected by dynamical backaction.

On the other hand, the eigenmode $\hat{B}=\frac{1}{\sqrt{N}}\sum_i \hat{b}_i$ with eigenvalue $\left\{-\Gamma+N\Gamma_\mathrm{dba}\right\}$ (the equivalent of the superradiant mode in cavity QED) is the only collective mode coupled to the cavity and its backaction damping rate is enhanced by a factor $N$, i.e. $\Gamma_\mathrm{dba}^{\left(B\right)}\equiv N\cdot\Gamma_\mathrm{dba}$. This scaling translates to a $\sqrt{N}$ scaling of $g_{\nu,0}$ (see equation 8 of the main text).

\subsection{Description of the Raman cross-section using the optomechanical formalism}
Writing the Stokes scattered power as {$P_{S}/\hbar\omega_L=\eta\kappa|\epsilon a_S|^2$} and expressing $a_S$ following eq. (\ref{Stokesfield}) yields
\begin{equation}
P_{S}/\hbar\omega_L=|\frac{1}{2}x_{\nu}G_{\nu}\left(\frac{\sqrt{\eta\kappa}}{-i\Delta +\kappa/2}\right)\left(\frac{s\sqrt{\eta\kappa}}{\left(\Omega_m-\Delta\right)-i\kappa/2}\right)|^2
\end{equation}
Here, we assume that the vibrational amplitude is related to the thermal energy, i.e. $\frac{1}{2}m_{\nu}\Omega_{\nu}^2\langle x_{\nu}^2\rangle=\frac{1}{2}k_B T$. 
We describe the amplitude of molecular vibrations with the help of its quantum mechanical description
\begin{equation}
x_{\nu}=\langle n_{\nu}+1|\hat{x}_{\nu}|n_{\nu}\rangle = \sqrt{\frac{\hbar}{2m_{\nu}\Omega_{\nu}}}\sqrt{n_{\nu}+1} = {x_\mathrm{zpm,\nu}\sqrt{n_{\nu}+1} }
\end{equation}
The Stokes cross-section is defined as $\sigma_{S}=\frac{P_{S}}{I}$ with $I=P_\mathrm{in}/A_\mathrm{eff}$ ($P_\mathrm{in}$ is the incident power {and $A_\mathrm{eff}$ the  illuminating spot area \footnote{Since the laser spot is large compared to the size of the plasmonic system the radiative coupling rate satisfies $\kappa_{ex}\propto A_\mathrm{eff}^{-1}$ so that 
$\kappa_{ex}A_\mathrm{eff}$ is approximately constant.}}). {As the incoming photon flux number is $|s|^{2}=P_\mathrm{in}/\hbar\omega_{L}$,} $\sigma_{S}$ is found to be
\begin{equation}
\sigma_{S}^{em}=\frac{\eta A_\mathrm{eff}}{4}\left(\frac{G_{\nu}x_\mathrm{zpm,\nu}}{\kappa}\right)^2\left(\frac{1}{\Delta^2/\kappa^2 + 1/4}\right)\left(\frac{1}{\left(\Delta-\Omega_m\right)^2/\kappa^2
+1/4}\right)\left(\bar{n}_{\nu}+1\right)
\end{equation}
where $\bar{n}_{\nu}$ is the average occupation number of the mode $\nu$ in equilibrium with an environment at temperature $T$. 
The same development can be done for the anti-Stokes process, leading to the result expressed in the main text.

%\subsection{Recovering the ``$E^4$ law"}
%
%The intuition that a more localized field leads to an electric field
%of greater amplitude in that region can be formalized as follows. We can use the expression
%$\frac{P_\mathrm{in}}{A_\mathrm{eff}}=I=\frac{1}{2}\epsilon_{0}c|E_{in}|^{2}$ to
%describe the incoming electric field and the  expression $P_\mathrm{in}\kappa^{-1}\sim\frac{1}{2}\epsilon_{0}|E_{loc}|^{2}V_{m}$ for the field inside the plasmonic cavity.
%The field enhancement factor $F^2=\frac{|E_{loc}|^{2}}{|E_{in}|^{2}}$
%can therefore be approximated by $\frac{cA_\mathrm{eff}}{\kappa V_{m}}$, where
%$\kappa$ is the total dissipation rate including both the metallic losses and the radiated power, and is in general frequency-dependent.

\section{Plasmon-vibration vacuum optomechanical coupling rate}

The change in resonance frequency of a cavity when a dielectric is inserted in an air gap is described by \cite{jackson} 
\begin{equation}
\Delta\omega_p=-\frac{\omega_{p}}{2}\frac{\int_{V}\vec{P}\cdot\vec{E_{p}}\ dV}{U_\mathrm{cav}}
\end{equation}
where $U_\mathrm{cav}=\frac{1}{2}\int\epsilon_{0}\frac{\mathrm{d}\left(\omega\epsilon\left(\omega\right)\right)}{\mathrm{d}\omega}|\vec{E_{p}}|^{2}dV$ is the energy stored inside the plasmonic cavity \cite{landau}, $\vec{E_{p}}$ the plasmonic cavity field and $\vec{P}$ the induced dipole per unit volume. {As long as the plasmonic cavity dimension is sufficiently small in comparison to the incoming wavelength, the quasistatic approximation remains valid and the magnetic energy can be neglected.} We can express the contribution of the electric field to the total energy stored in the metal by $U_\mathrm{cav}=\mu U_d$ where $U_d=\frac{1}{2}\int\epsilon_{0}|\vec{E_{p}}|^{2}dV$ and the factor $\mu$ depends on the dielectric function of the metal and the plasmon resonance frequency \cite{wang}. 

We assume that a single molecule is located at the position of maximum electric
field. Introducing the molecular dipole moment $\vec{p}$
\begin{equation}
\Delta\omega_p\approx-\frac{\omega_{p}}{2}\frac{\vec{p}\cdot\vec{E}_\mathrm{max}}{\mu U_{d}}
\end{equation}
The induced dipole is related to the electric field by $\vec{p}=\alpha\cdot\vec{E}_\mathrm{max}$; the dependence of the linear polarizability $\alpha$ on the molecular displacement $x_{\nu}$ is developed to first order 
\begin{equation}
\vec{p}=\alpha_{}(x_{\nu})\cdot\vec{E}_\mathrm{max}\approx\left(\alpha_{}(0)+\frac{\partial \alpha_{}}{\partial x_{\nu}}x_{\nu}\right)\cdot\vec{E}_\mathrm{max}
\end{equation}
where the gradient of polarizability $\frac{\partial \alpha}{\partial x_\nu}$ has units $\left[\epsilon_{0}\mathrm{m}^{2}\right].$
Note that in the expression for the polarizability of the molecule contains the contributions from the internal electonic transitions so that it is frequency dependent.

We can now quantify the sensitivity of the plasmonic frequency to the molecular vibrations to first order in $x_\nu$
\begin{equation}
G_{\nu}=\frac{\partial\left(\Delta\omega_{p}\right)}{\partial x_{\nu}}=-\frac{\omega_{p}}{2}\frac{\frac{\partial\alpha_{}}{\partial x_{\nu}}|\vec{E}_\mathrm{max}|^{2}}{\mu U_{d}}=-\frac{\omega_{p}}{\epsilon_{0}V_{m}}\frac{\partial\alpha_{}}{\partial x_{\nu}}
\end{equation}
expressed as a function of the effective mode volume of the cavity $V_m\doteq \frac{\mu U_{d}}{\frac{1}{2}\epsilon_0|E_\mathrm{max}|^2}$. When the plasmon resonance frequency is getting closer to the bulk plasma frequency, the energy stored in the metal grows and the mode volume defined here is getting larger \cite{wang}.

We have thus shown that the plasmon frequency is coupled to the molecular
vibration and that this coupling can be quantified by the optomechanical
vacuum coupling rate (also called single photon coupling rate) 
\begin{equation}
g_{\nu,0}=G_{\nu}x_\mathrm{zpm,\nu}=\omega_{p}\frac{\partial\alpha}{\partial x_{\nu}}\frac{1}{V_{m}\epsilon_{0}}\sqrt{\frac{\hbar}{2m_{\nu}\Omega_{\nu}}}\label{eq:g0th}
\end{equation}
Applying the concepts used in optomechanics, we are thus able to find
a quantitative way of describing the interaction between the molecular
vibrations and the plasmonic resonance. 

\section{Quantifying the optomechanical coupling}

\subsection{Calculation of $g_{0}$}

Following the conventions used in Raman spectroscopy we use instead of the normal coordinates $x_{\nu}$ the reduced coordinates
$Q_{\nu}$ with units $\left[\mathrm{kg}^{-1/2}\mathrm{m}\right]$,
refered to as \textit{the mass-weighted cartesian displacement coordinates} \cite{wilson}.
In this notation the kinetic energy is given by $E_K=\frac{1}{2}\sum_{\nu=1}^{3N}\dot{Q}_{\nu}^{2}$.
Expressing the vacuum coupling rate (\ref{eq:g0th}) in the reduced
coordinates ($Q_{\nu}=\sqrt{m_\mathrm{eff,\nu}}x_{\nu}$) leads to
\begin{equation}
g_{\nu,0}=\omega_{p}\left(\frac{\partial\alpha}{\partial Q_{\nu}}\right)\left(\frac{1}{V_{m}\epsilon_{0}}\right)\sqrt{\frac{\hbar}{2\Omega_{\nu}}}
\end{equation}
In optomechanical systems the effective mass is defined to verify the equipartition theorem ($\frac{1}{2}m_\mathrm{eff,\nu}\Omega_{\nu}^{2}x_{\nu}^{2}=U_{\nu}$),
with $U_{\nu}$ the energy stored in this mode. 

In order to compute the optomechanical vacuum coupling rate, we need the value of the Raman polarizability tensor. The Raman activity can be inferred from experimental data and, using density functional theory (DFT) simulations, it is possible to have access to the tensor
components. Generally, the Raman intensity, describing the orientation-averaged magnitude of the Raman scattering, is given by two invariant scalars built from the space derivative of the polarizability tensor 
\begin{equation}
R_{\nu}=\frac{45\bar{\alpha}_{\nu}^{2}+7\bar{\gamma}_{\nu}^{2}}{45}
\end{equation}
where $\bar{\alpha}_{\nu}^{2}$ and $\bar{\gamma}_{\nu}^{2}$ are
the isotropic and anisotropic parts of the tensor.

In a simplified one dimensional model, the Raman activity can
be described as $R_{\nu}=\left(\frac{\partial\alpha}{\partial Q_{\nu}}\right)^{2}$
with units $\left[\epsilon_{0}^{2}\mathrm{m}^{4}\mathrm{kg}^{-1}\right]$.
In simulations, using gaussian units, this first quantity is however
expressed in $\left[\mathrm{\mathring{A}}^{4}\mathrm{amu}^{-1}\right]$.
The relation between the usual description of the Raman activity and
that one is 
\begin{equation}
R_{\nu}[\text{SI}]=(4\pi\epsilon_{0})^{2}\left[\frac{10^{-40}}{1.66\cdot10^{-27}}\right]R_{\nu}[\text{g.u.}]
\end{equation}

Consequently, the vacuum optomechanical coupling rate (\ref{eq:g0th})
is given by 
\begin{equation}
g_{0}=\frac{4\pi\epsilon_{0}\cdot10^{-20}}{\sqrt{1.66\cdot10^{-27}}}\omega_{p}\left[\frac{1}{V_{m}\epsilon_{0}}\right]\sqrt{\frac{\hbar}{2\Omega_{\nu}}}\sqrt{R_{\nu}[\text{g.u.}]}
\end{equation}

Experimental studies are usually expressing the resonance of the plasmon
in {[}nm{]} and the wavenumber $\tilde{\nu}$ of the vibrational mode in {[}cm$^{-1}${]},
being related to the angular frequency via 
\begin{equation}
\Omega_{\nu}\ [\mathrm{Hz}]=2\pi c\tilde{\nu}[\text{cm}^{-1}]\cdot10^{2}
\end{equation}
For numerical approximations below we will consider that the resonance of the plasmon is around 900~nm and we will express
it in angular frequency  $\omega_{p}\ [\mathrm{Hz}]=\frac{2\pi c}{\lambda_{p}[\text{nm}]}\cdot10^{9}$.

The vacuum coupling rate can finally be expressed as a function of experimentally used parameters 
\begin{equation}
g_{0}\ [\mathrm{Hz}]=\left(\frac{1}{V_{m}}\right)4\cdot10^{\frac{3}{2}}\sqrt{\frac{c\hbar\pi^{3}}{1.66}}\frac{\sqrt{R_{\nu}[\mathrm{g.u.}]}}{\lambda_{p}[\text{nm}]\sqrt{\tilde{\nu}[\text{cm}^{-1}]}}
\end{equation}

\subsection{Examples of coupling rate values}

We computed the vacuum optomechanical coupling rate between several molecules and the localized surface plasmon of a dimer nanoparticle. Molecular parameters were obtained from the literature. We used experimental parameters \cite{Zhu_2014} to give a realistic approximation of the mode volume. According to the geometrical estimate used in \cite{Savage}, we can approximate the mode volume arising from the field contribution only as $V_m^d\sim r^{1/2}d^{3/2}h$ with $r$ the radius of the cylinder, $h$ its height and $d$ the size of the gap. We note that for the experimental case studied { the plasmonic resonance frequency is sufficiently below the bulk plasma frequency of gold so that} the contributions from the field and charges can be considered equals \cite{wang}. {We recover in that case the expected equivalence between the two energy contributions valid generally for non-dispersive materials and the effective mode volume is then given by} $V_m=\mu V_m^d$ with $\mu\simeq2$. We present in Table \ref{tab:g0} the molecular parameters of interest and the related optomechanical coupling for different vibrational modes.
 
%The plasmonic field of a silver dimer was simulated with the finite element modeling (FEM) software \textsc{Comsol} using the Drude model to describe the nanoparticles (Section 3 in the Supplementary Material). For a model system consisting of two 10~nm radius silver spheres separated by a 2~nm gap we find a mode volume $V_m\simeq1.13\cdot10^{-6}\ \lambda^{3}$, compatible with values inferred from previous works \cite{leru2006c}. We present in Table \ref{tab:g0} the molecular parameters of interest and the related optomechanical coupling for different vibrational modes.

\begin{centering}

\begin{table}[h]
\begin{tabular}{|c|c|c|c|}
\hline 
molecule & $\Omega_{\nu}/2\pi${[}THz{]}({[}cm$^{-1}${]}) & $R_{\nu}${[}\AA$^{4}$amu$^{-1}${]} & estimated $g_{\nu,0}/2\pi${[}Hz{]}\tabularnewline
\hline 
\hline 
R6G\cite{watanabe} & 39 (1301) & 5.9 & 7.9$\cdot$10$^{9}$\tabularnewline
\hline 
R6G & 40.5 (1351) & 351.7 & 6.0$\cdot$10$^{10}$\tabularnewline
\hline 
Thiophenol\cite{humbert} & 29.9 (998) & 31.6 & 2.1$\cdot$10$^{10}$\tabularnewline
\hline 
Thiophenol & 32.1 (1072) & 1.7 & 4.7$\cdot$10$^{9}$\tabularnewline
\hline 
GBT\cite{humbert} & 30 (1000) & 29.5 & 2.0$\cdot$10$^{10}$\tabularnewline
\hline 
GBT & 32.2 (1075) & 388.5 & 7.0$\cdot$10$^{10}$\tabularnewline
\hline 
$G$-band of graphene or CNT & $\sim$48 ($\sim 1600$) & $\sim 10^3$--10$^4$ & $\sim 10^{10}$--10$^{11}$\tabularnewline
\hline 
\end{tabular}
\protect\caption{Vacuum optomechanical coupling rates ($g_{\nu,0}/2\pi$) calculated with parameters from the literature. We used a plasmonic mode volume  $V_{m}=2.02\cdot 10^{-7}\lambda^{3}$ and resonance of the plasmon $\lambda_{p}=900$~nm (corresponding to the experimantal case \cite{Zhu_2014}). Values for graphene and carbon nanotubes (CNT) are estimated using Raman activity from the literature \cite{Narula_2010} 
(we obtain similar values for the radial breathing mode of CNTs). \label{tab:g0}}
\end{table}
\end{centering}

\begin{figure}[h!]
\begin{centering}
\vspace{-0.7in}
\includegraphics[scale=0.47]{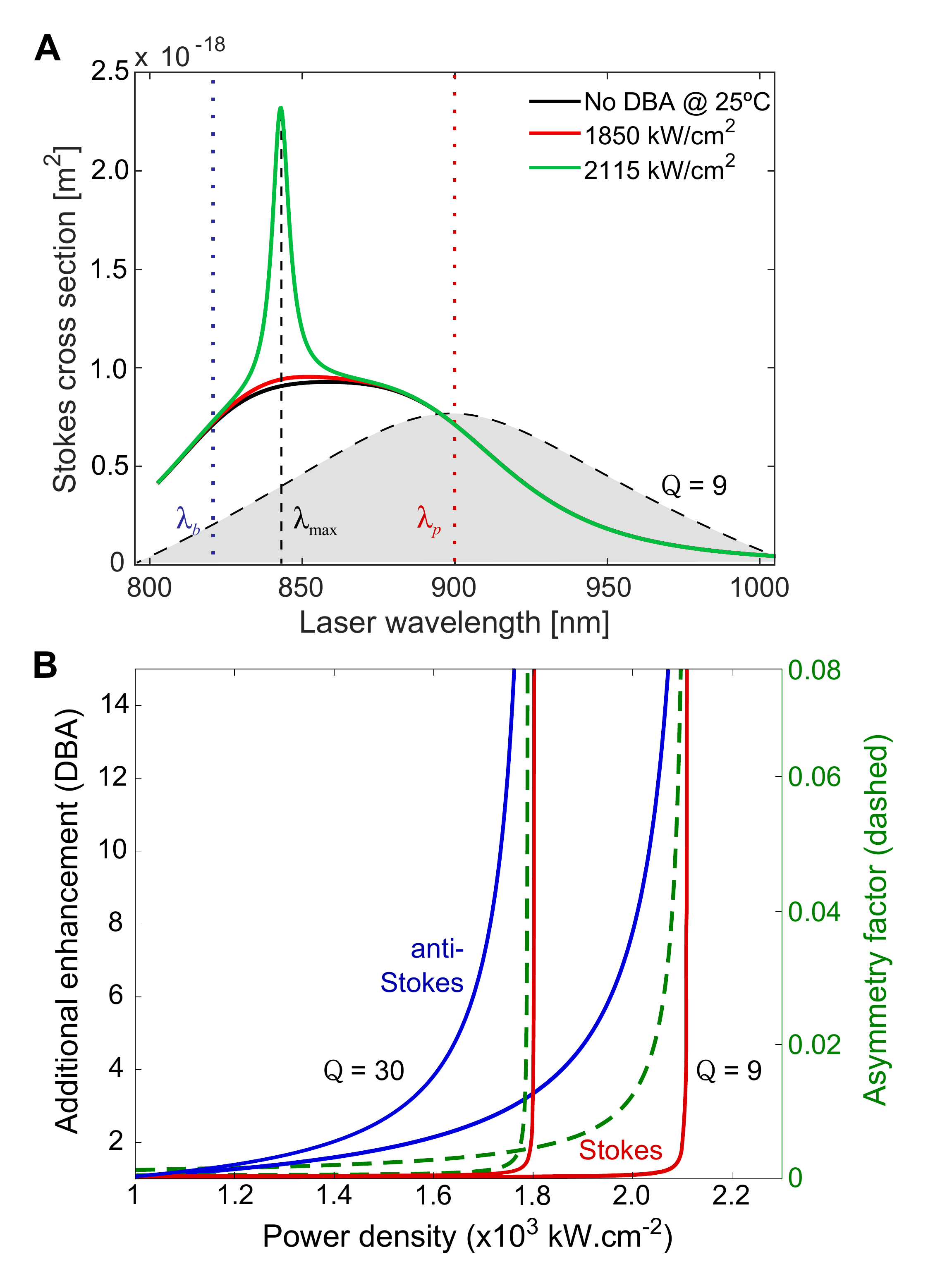}
\par\end{centering}
\protect\caption{\label{fig:EM}
\textbf{Plasmon-enhanced cross-section of a single GBT molecule.} 
(\textbf{A}) Computed Stokes cross-section as a function of laser wavelength for different powers close to threshold. The cross-section calculated without dynamical backaction amplification (No DBA) is shown in black and corresponds to the conventional field enhancement model. The shaded area shows the lineshape (arbitrary scale) of the plasmonic resonance ($Q=9$). We assume here a mode volume $V_m\simeq3.4\cdot10^{-8}\lambda^{3}$ \cite{Savage} for which $g_{\nu,0}/2\pi = 4.2\cdot 10^{11}$~Hz. { The plasmonic resonance is at $\lambda_p$ and the blue phonon sideband at $\lambda_b$.}
(\textbf{B}) Additional enhancement factor due to dynamical backaction amplification for the Stokes (red) and anti-Stokes (blue) cross-section under excitation at the wavelength $\lambda_\mathrm{max}=843$~nm marked in (\textbf{A}). For narrower plasmonic resonance ($Q=30$)  the instability threshold is lowered and occurs under excitation closer to the phonon sideband ($\lambda_\mathrm{max}=823$~nm). { For these two cases the anti-Stokes/Stokes ratio (green dashed line, right scale) is also given for completeness.}
}
\end{figure}

\section{Single molecule scheme}

As an example of single molecule SERS, we consider the 1075~cm$^{-1}$ vibrational mode of GBT {(gold-benzenethiolate)} coupled to the realistic experimental dimer considered in the main text. Its single-photon cooperativity is {$C_{0}=C/\bar{n}_{p}$} = $8.20\cdot 10^{-3}=O(10^{-2})$. Under maximal realistic pump intensity $I\sim 10$~MW$\cdot$cm$^{-2}$ we find that the photon number inside the plasmon cavity  is $\bar{n}_{p}\sim 13.2$ and thus $C=0.11$, still below threshold.  Assuming that the mode volume of the ``hottest" spots reaches the quantum limit $V_m\simeq3.4\cdot10^{-8}\lambda^{3}$ \cite{Savage}, the cooperativity becomes unity for a pump intensity of only $I=2.1$~MW$\cdot$cm$^{-2}$ (Fig.~\ref{fig:EM}), i.e. with less than five plasmons in the cavity on average (we therefore do not expect nonlinear mechanisms to be significant in the plasmon). For a narrower plasmon resonance with $Q=30$ we find the threshold intensity to be around 1.8~MW$\cdot$cm$^{-2}$ (Fig.~\ref{fig:EM}B).

\section{Nonlinearity and super-resolution}

\begin{figure}[H]
\centering{}\includegraphics[scale=0.7]{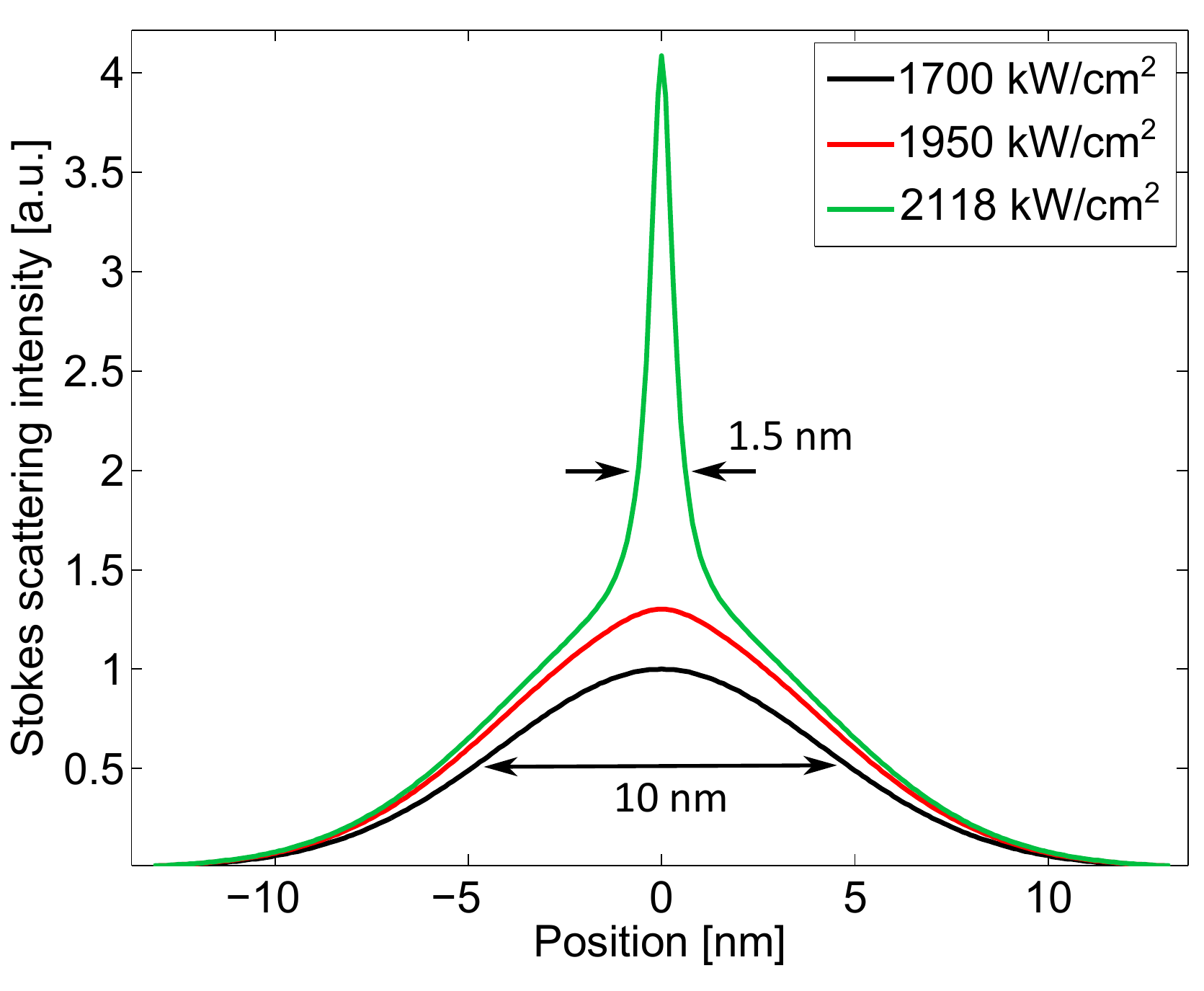}
\protect\caption{
\textbf{Super-resolution in TERS} 
Computed Stokes scattering intensity as a function of lateral position assuming a point-source scatterer located at the origin and a Gaussian plasmonic field distribution of 10~nm full-width at half maximum. All parameters are as in Fig. \ref{fig:EM} (keeping the GBT molecule as an example). Close to the instability threshold (green curve) the pronounced nonlinearity of the response leads to a sharp increase in spatial resolution. \label{fig:superres}}
\end{figure}

The recent experiments of Zhang et al. \cite{Zhang2013,Jiang_2015} demonstrating sub-nanometer resolution in TERS have spurred excitement and debates over the physical mechanism leading to super-resolution, since the plasmonic hot spot lateral size at the tip apex cannot be smaller than $\sim 10$~nm. It is clear that strong nonlinear effects must be involved. Interestingly, in similar excitation conditions as in \cite{Zhang2013} (blue detuned, close to the phonon sideband), our model predicts a large nonlinear response when working close to the threshold of parametric instability. In Fig.~\ref{fig:superres} we exemplify how our model can lead to super-resolution TERS imaging.


\begin{thebibliography}{10}
\expandafter\ifx\csname url\endcsname\relax
  \def\url#1{\texttt{#1}}\fi
\expandafter\ifx\csname urlprefix\endcsname\relax\def\urlprefix{URL }\fi
\providecommand{\bibinfo}[2]{#2}
\providecommand{\eprint}[2][]{\url{#2}}

\bibitem{Fleischmann_1974}
\bibinfo{author}{Fleischmann, M.}, \bibinfo{author}{Hendra, P.} \&
  \bibinfo{author}{McQuillan, A.}
\newblock \bibinfo{title}{Raman spectra of pyridine adsorbed at a silver
  electrode}.
\newblock \emph{\bibinfo{journal}{Chemical Physics Letters}}
  \textbf{\bibinfo{volume}{26}}, \bibinfo{pages}{163 -- 166}
\newblock  (\bibinfo{year}{1974}).

\bibitem{VanDuyne1977}
\bibinfo{author}{Jeanmaire, D.~L.} \& \bibinfo{author}{Duyne, R. P.~V.}
\newblock \bibinfo{title}{Surface raman spectroelectrochemistry: Part i.
  heterocyclic, aromatic, and aliphatic amines adsorbed on the anodized silver
  electrode}.
\newblock \emph{\bibinfo{journal}{Journal of Electroanalytical Chemistry and
  Interfacial Electrochemistry}} \textbf{\bibinfo{volume}{84}},
  \bibinfo{pages}{1 -- 20}
\newblock  (\bibinfo{year}{1977}).

\bibitem{Kneipp1997}
\bibinfo{author}{Kneipp, K.} \emph{et~al.}
\newblock \bibinfo{title}{Single molecule detection using surface-enhanced
  raman scattering (sers)}.
\newblock \emph{\bibinfo{journal}{Phys. Rev. Lett.}}
  \textbf{\bibinfo{volume}{78}}, \bibinfo{pages}{1667--1670}
\newblock  (\bibinfo{year}{1997}).

\bibitem{NieScience}
\bibinfo{author}{Nie, S.}
\newblock \bibinfo{title}{Probing single molecules and single nanoparticles by
  surface-enhanced raman scattering}.
\newblock \emph{\bibinfo{journal}{Science}} \textbf{\bibinfo{volume}{275}},
  \bibinfo{pages}{1102--1106}
\newblock  (\bibinfo{year}{1997}).

\bibitem{Shalaev}
\bibinfo{author}{Shalaev, V.} \& \bibinfo{author}{Sarychev, A.}
\newblock \bibinfo{title}{Nonlinear optics of random metal-dielectric films}.
\newblock \emph{\bibinfo{journal}{Phys. Rev. B}} \textbf{\bibinfo{volume}{57}}
\newblock  (\bibinfo{year}{1998}).

\bibitem{Pettinger}
\bibinfo{author}{Pettinger, B.}, \bibinfo{author}{Schambach, P.},
  \bibinfo{author}{Villagomez, C.~J.} \& \bibinfo{author}{Scott, N.}
\newblock \bibinfo{title}{Tip-enhanced raman spectroscopy: near-fields acting
  on a few molecules}.
\newblock \emph{\bibinfo{journal}{Annual Reviews of Physical Chemistry}}
  \textbf{\bibinfo{volume}{63}}, \bibinfo{pages}{379--99}
\newblock  (\bibinfo{year}{2012}).

\bibitem{Sharma}
\bibinfo{author}{Sharma, B.} \emph{et~al.}
\newblock \bibinfo{title}{High-performance sers substrates: Advances and
  challenges}.
\newblock \emph{\bibinfo{journal}{MRS Bulletin}} \textbf{\bibinfo{volume}{38}},
  \bibinfo{pages}{615--624}
\newblock  (\bibinfo{year}{2013}).

\bibitem{Qian}
\bibinfo{author}{Qian, X.~M.} \& \bibinfo{author}{Nie, S.~M.}
\newblock \bibinfo{title}{Single-molecule and single-nanoparticle sers: from
  fundamental mechanisms to biomedical applications}.
\newblock \emph{\bibinfo{journal}{Chem Soc Rev}} \textbf{\bibinfo{volume}{37}},
  \bibinfo{pages}{912--20}
\newblock  (\bibinfo{year}{2008}).

\bibitem{Pendry}
\bibinfo{author}{Luo, Y.}, \bibinfo{author}{Aubry, A.} \&
  \bibinfo{author}{Pendry, J.~B.}
\newblock \bibinfo{title}{Electromagnetic contribution to surface-enhanced
  raman scattering from rough metal surfaces: A transformation optics
  approach}.
\newblock \emph{\bibinfo{journal}{Phys. Rev. B}} \textbf{\bibinfo{volume}{83}},
  \bibinfo{pages}{155422}
\newblock  (\bibinfo{year}{2011}).

\bibitem{Maher2004}
\bibinfo{author}{Maher, R.~C.} \emph{et~al.}
\newblock \bibinfo{title}{Stokes/anti-stokes anomalies under surface enhanced
  raman scattering conditions}.
\newblock \emph{\bibinfo{journal}{The Journal of Chemical Physics}}
  \textbf{\bibinfo{volume}{120}}, \bibinfo{pages}{11746}
\newblock  (\bibinfo{year}{2004}).

\bibitem{LeRu2006b}
\bibinfo{author}{Le~Ru, E.~C.} \& \bibinfo{author}{Etchegoin, P.~G.}
\newblock \bibinfo{title}{Vibrational pumping and heating under sers
  conditions: fact or myth?}
\newblock \emph{\bibinfo{journal}{Faraday Discussions}}
  \textbf{\bibinfo{volume}{132}}, \bibinfo{pages}{63}
\newblock  (\bibinfo{year}{2006}).

\bibitem{Zhang2013}
\bibinfo{author}{Zhang, R.} \emph{et~al.}
\newblock \bibinfo{title}{Chemical mapping of a single molecule by
  plasmon-enhanced raman scattering}.
\newblock \emph{\bibinfo{journal}{Nature}} \textbf{\bibinfo{volume}{498}},
  \bibinfo{pages}{82--6}
\newblock  (\bibinfo{year}{2013}).

\bibitem{Zhu_2014}
\bibinfo{author}{Zhu, W.} \& \bibinfo{author}{Crozier, K.~B.}
\newblock \bibinfo{title}{Quantum mechanical limit to plasmonic enhancement as
  observed by surface-enhanced raman scattering}.
\newblock \emph{\bibinfo{journal}{Nat Commun}} \textbf{\bibinfo{volume}{5}}
\newblock  (\bibinfo{year}{2014}).

\bibitem{Jiang_2015}
\bibinfo{author}{Jiang, S.} \emph{et~al.}
\newblock \bibinfo{title}{Distinguishing adjacent molecules on a surface using
  plasmon-enhanced raman scattering}.
\newblock \emph{\bibinfo{journal}{Nat Nano}} \emph{\bibinfo{volume}{advance
  online publication}}
\newblock  (\bibinfo{year}{2015}).

\bibitem{atkin}
\bibinfo{author}{Atkin, J.~M.} \& \bibinfo{author}{Raschke, M.~B.}
\newblock \bibinfo{title}{Techniques: Optical spectroscopy goes
  intramolecular}.
\newblock \emph{\bibinfo{journal}{Nature}} \textbf{\bibinfo{volume}{498}},
  \bibinfo{pages}{44--45}
\newblock  (\bibinfo{year}{2013}).

\bibitem{Kippenberg2008}
\bibinfo{author}{Kippenberg, T.~J.} \& \bibinfo{author}{Vahala, K.~J.}
\newblock \bibinfo{title}{Cavity {O}ptomechanics: {B}ack-{A}ction at the
  {M}esoscale}.
\newblock \emph{\bibinfo{journal}{Science}} \textbf{\bibinfo{volume}{321}},
  \bibinfo{pages}{1172--1176}
\newblock  (\bibinfo{year}{2008}).

\bibitem{Braginsky1967}
\bibinfo{author}{Braginsky, V.} \& \bibinfo{author}{Manukin, A.}
\newblock \bibinfo{title}{Ponderomotive effects of electromagnetic radiation}.
\newblock \emph{\bibinfo{journal}{Soviet Physics JETP}}
  \textbf{\bibinfo{volume}{25}}, \bibinfo{pages}{653}
\newblock  (\bibinfo{year}{1967}).

\bibitem{Kippenberg2005}
\bibinfo{author}{Kippenberg, T.}, \bibinfo{author}{Rokhsari, H.},
  \bibinfo{author}{Carmon, T.}, \bibinfo{author}{Scherer, A.} \&
  \bibinfo{author}{Vahala, K.}
\newblock \bibinfo{title}{Analysis of radiation-pressure induced mechanical
  oscillation of an optical microcavity}.
\newblock \emph{\bibinfo{journal}{Phys.\ Rev.\ Lett.}}
  \textbf{\bibinfo{volume}{95}}, \bibinfo{pages}{033901}
\newblock  (\bibinfo{year}{2005}).

\bibitem{Aspelmeyer_2013}
\bibinfo{author}{Aspelmeyer, M.}, \bibinfo{author}{Kippenberg, T.~J.} \&
  \bibinfo{author}{Marquardt, F.}
\newblock \bibinfo{title}{Cavity optomechanics}.
\newblock \emph{\bibinfo{journal}{Reviews of Modern Physics}}
  \textbf{\bibinfo{volume}{86}}, \bibinfo{pages}{1391}
\newblock  (\bibinfo{year}{2014}).

\bibitem{Gorodetsky}
\bibinfo{author}{Gorodetsky, M.~L.}, \bibinfo{author}{Schliesser, A.},
  \bibinfo{author}{Anetsberger, G.}, \bibinfo{author}{Deleglise, S.} \&
  \bibinfo{author}{Kippenberg, T.~J.}
\newblock \bibinfo{title}{Determination of the vacuum optomechanical coupling
  rate using frequency noise calibration}.
\newblock \emph{\bibinfo{journal}{Opt. Express}} \textbf{\bibinfo{volume}{18}},
  \bibinfo{pages}{23236--23246}
\newblock  (\bibinfo{year}{2010}).

\bibitem{vanLaer2015}
\bibinfo{author}{Van~Laer, R.}, \bibinfo{author}{Kuyken, B.},
  \bibinfo{author}{Baets, R.} \& \bibinfo{author}{Van~Thourhout, D.}
\newblock \bibinfo{title}{Unifying brillouin scattering and cavity
  optomechanics}.
\newblock \emph{\bibinfo{journal}{arXiv preprint arXiv:1503.03044}}
\newblock  (\bibinfo{year}{2015}).

\bibitem{ebbesen}
\bibinfo{author}{Shalabney, A.} \emph{et~al.}
\newblock \bibinfo{title}{Coherent coupling of molecular resonators with a
  microcavity mode}.
\newblock \emph{\bibinfo{journal}{Nature communications}}
  \textbf{\bibinfo{volume}{6}}
\newblock  (\bibinfo{year}{2015}).

\bibitem{Long_2015}
\bibinfo{author}{Long, J.~P.} \& \bibinfo{author}{Simpkins, B.~S.}
\newblock \bibinfo{title}{Coherent coupling between a molecular vibration and
  fabry--perot optical cavity to give hybridized states in the strong coupling
  limit}.
\newblock \emph{\bibinfo{journal}{ACS Photonics}} \textbf{\bibinfo{volume}{2}},
  \bibinfo{pages}{130--136}
\newblock  (\bibinfo{year}{2015}).

\bibitem{koenderink}
\bibinfo{author}{Koenderink, A.~F.}
\newblock \bibinfo{title}{On the use of purcell factors for plasmon antennas}.
\newblock \emph{\bibinfo{journal}{Opt. Lett.}} \textbf{\bibinfo{volume}{35}},
  \bibinfo{pages}{4208--4210}
\newblock  (\bibinfo{year}{2010}).

\bibitem{Botter_2012}
\bibinfo{author}{Botter, T.}, \bibinfo{author}{Brooks, D. W.~C.},
  \bibinfo{author}{Brahms, N.}, \bibinfo{author}{Schreppler, S.} \&
  \bibinfo{author}{Stamper-Kurn, D.~M.}
\newblock \bibinfo{title}{Linear amplifier model for optomechanical systems}.
\newblock \emph{\bibinfo{journal}{Phys. Rev. A}} \textbf{\bibinfo{volume}{85}},
  \bibinfo{pages}{013812}
\newblock  (\bibinfo{year}{2012}).

\bibitem{wilson}
\bibinfo{author}{Wilson, E.}, \bibinfo{author}{Decius, J.} \&
  \bibinfo{author}{Cross, P.}
\newblock \emph{\bibinfo{title}{Molecular Vibrations: The Theory of Infrared
  and Raman Vibrational Spectra}}
\newblock  (\bibinfo{publisher}{Dover Publications}, \bibinfo{year}{1955}).

\bibitem{Chan_2012}
\bibinfo{author}{Chan, J.}, \bibinfo{author}{Safavi-Naeini, A.~H.},
  \bibinfo{author}{Hill, J.~T.}, \bibinfo{author}{Meenehan, S.} \&
  \bibinfo{author}{Painter, O.}
\newblock \bibinfo{title}{Optimized optomechanical crystal cavity with acoustic
  radiation shield}.
\newblock \emph{\bibinfo{journal}{Applied Physics Letters}}
  \textbf{\bibinfo{volume}{101}}, \bibinfo{pages}{--}
\newblock  (\bibinfo{year}{2012}).

\bibitem{LeRuCh6}
\bibinfo{author}{Ru, E. C.~L.} \& \bibinfo{author}{Etchegoin, P.~G.}
\newblock In \emph{\bibinfo{booktitle}{Principles of Surface-Enhanced Raman
  Spectroscopy}}, \bibinfo{pages}{299 -- 365 (Ch. 6)}
\newblock  (\bibinfo{publisher}{Elsevier}, \bibinfo{address}{Amsterdam},
  \bibinfo{year}{2009}).

\bibitem{Nordlander2011}
\bibinfo{author}{Zuloaga, J.} \& \bibinfo{author}{Nordlander, P.}
\newblock \bibinfo{title}{On the energy shift between near-field and far-field
  peak intensities in localized plasmon systems}.
\newblock \emph{\bibinfo{journal}{Nano Letters}} \textbf{\bibinfo{volume}{11}},
  \bibinfo{pages}{1280--1283}
\newblock  (\bibinfo{year}{2011}).

\bibitem{Moreno2013}
\bibinfo{author}{Moreno, F.}, \bibinfo{author}{Albella, P.} \&
  \bibinfo{author}{Nieto-Vesperinas, M.}
\newblock \bibinfo{title}{{Analysis of the spectral behavior of localized
  plasmon resonances in the near- and far-field regimes}}.
\newblock \emph{\bibinfo{journal}{Langmuir}} \textbf{\bibinfo{volume}{29}},
  \bibinfo{pages}{6715--6721}
\newblock  (\bibinfo{year}{2013}).

\bibitem{Ruben2015}
\bibinfo{author}{Esteban, R.} \emph{et~al.}
\newblock \bibinfo{title}{The morphology of narrow gaps modifies the plasmonic
  response}.
\newblock \emph{\bibinfo{journal}{ACS Photonics}} \textbf{\bibinfo{volume}{2}},
  \bibinfo{pages}{295--305}
\newblock  (\bibinfo{year}{2015}).

\bibitem{boyd}
\bibinfo{author}{Boyd, R.}
\newblock \emph{\bibinfo{title}{Nonlinear Optics}}
\newblock  (\bibinfo{publisher}{Elsevier Science}, \bibinfo{year}{2003}).

\bibitem{camden}
\bibinfo{author}{Camden, J.~P.} \emph{et~al.}
\newblock \bibinfo{title}{Probing the structure of single-molecule
  surface-enhanced raman scattering hot spots}.
\newblock \emph{\bibinfo{journal}{Journal of the American Chemical Society}}
  \textbf{\bibinfo{volume}{130}}, \bibinfo{pages}{12616--12617}
\newblock  (\bibinfo{year}{2008}).

\bibitem{Schliesser_2006}
\bibinfo{author}{Schliesser, A.}, \bibinfo{author}{Del'Haye, P.},
  \bibinfo{author}{Nooshi, N.}, \bibinfo{author}{Vahala, K.~J.} \&
  \bibinfo{author}{Kippenberg, T.~J.}
\newblock \bibinfo{title}{Radiation pressure cooling of a micromechanical
  oscillator using dynamical backaction}.
\newblock \emph{\bibinfo{journal}{Phys. Rev. Lett.}}
  \textbf{\bibinfo{volume}{97}}, \bibinfo{pages}{243905}
\newblock  (\bibinfo{year}{2006}).

\bibitem{kenkre}
\bibinfo{author}{Kenkre, V.}, \bibinfo{author}{Tokmakoff, A.} \&
  \bibinfo{author}{Fayer, M.}
\newblock \bibinfo{title}{{Theory of vibrational relaxation of polyatomic
  molecules in liquids}}.
\newblock \emph{\bibinfo{journal}{Journal of Physical Chemistry}}
  \textbf{\bibinfo{volume}{101}}, \bibinfo{pages}{10618--10629}
\newblock  (\bibinfo{year}{1994}).

\bibitem{Savage}
\bibinfo{author}{Savage, K.~J.}, \bibinfo{author}{Hawkeye, M.~M.},
  \bibinfo{author}{Esteban, R.}, \bibinfo{author}{Borisov, J., Andrei
  G.and~Aizpurua} \& \bibinfo{author}{Baumberg, J.~J.}
\newblock \bibinfo{title}{Revealing the quantum regime in tunnelling
  plasmonics}.
\newblock \emph{\bibinfo{journal}{Nature}} \textbf{\bibinfo{volume}{491}}
\newblock  (\bibinfo{year}{2012}).

\bibitem{scholl2013observation}
\bibinfo{author}{Scholl, J.~A.}, \bibinfo{author}{Garc{\'\i}a-Etxarri, A.},
  \bibinfo{author}{Koh, A.~L.} \& \bibinfo{author}{Dionne, J.~A.}
\newblock \bibinfo{title}{Observation of quantum tunneling between two
  plasmonic nanoparticles}.
\newblock \emph{\bibinfo{journal}{Nano letters}} \textbf{\bibinfo{volume}{13}},
  \bibinfo{pages}{564--569}
\newblock  (\bibinfo{year}{2013}).

\bibitem{kipf}
\bibinfo{author}{Kipf, T.} \& \bibinfo{author}{Agarwal, G.~S.}
\newblock \bibinfo{title}{Superradiance and collective gain in multimode
  optomechanics}.
\newblock \emph{\bibinfo{journal}{Phys. Rev. A}} \textbf{\bibinfo{volume}{90}},
  \bibinfo{pages}{053808}
\newblock  (\bibinfo{year}{2014}).

\bibitem{giessen}
\bibinfo{author}{Liu, N.} \emph{et~al.}
\newblock \bibinfo{title}{Plasmonic analogue of electromagnetically induced
  transparency at the drude damping limit}.
\newblock \emph{\bibinfo{journal}{Nature materials}}
  \textbf{\bibinfo{volume}{8}}, \bibinfo{pages}{758--762}
\newblock  (\bibinfo{year}{2009}).

\bibitem{McFarland_2005}
\bibinfo{author}{McFarland, A.~D.}, \bibinfo{author}{Young, M.~A.},
  \bibinfo{author}{Dieringer, J.~A.} \& \bibinfo{author}{Van~Duyne, R.~P.}
\newblock \bibinfo{title}{Wavelength-scanned surface-enhanced raman excitation
  spectroscopy}.
\newblock \emph{\bibinfo{journal}{The Journal of Physical Chemistry B}}
  \textbf{\bibinfo{volume}{109}}, \bibinfo{pages}{11279--11285}
\newblock  (\bibinfo{year}{2005}).

\bibitem{Galland_2014}
\bibinfo{author}{Galland, C.}, \bibinfo{author}{Sangouard, N.},
  \bibinfo{author}{Piro, N.}, \bibinfo{author}{Gisin, N.} \&
  \bibinfo{author}{Kippenberg, T.~J.}
\newblock \bibinfo{title}{Heralded single-phonon preparation, storage, and
  readout in cavity optomechanics}.
\newblock \emph{\bibinfo{journal}{Phys. Rev. Lett.}}
  \textbf{\bibinfo{volume}{112}}, \bibinfo{pages}{143602}
\newblock  (\bibinfo{year}{2014}).

\end{thebibliography}

\begin{thebibliography}{10}
\expandafter\ifx\csname url\endcsname\relax
  \def\url#1{\texttt{#1}}\fi
\expandafter\ifx\csname urlprefix\endcsname\relax\def\urlprefix{URL }\fi
\providecommand{\bibinfo}[2]{#2}
\providecommand{\eprint}[2][]{\url{#2}}

\bibitem{Aspelmeyer_2013}
\bibinfo{author}{Aspelmeyer, M.}, \bibinfo{author}{Kippenberg, T.~J.} \&
  \bibinfo{author}{Marquardt, F.}
\newblock \bibinfo{title}{Cavity optomechanics}.
\newblock \emph{\bibinfo{journal}{Reviews of Modern Physics}}
  \textbf{\bibinfo{volume}{86}}, \bibinfo{pages}{1391}
\newblock  (\bibinfo{year}{2014}).

\bibitem{Kippenberg_07}
\bibinfo{author}{Kippenberg, T.~J.} \& \bibinfo{author}{Vahala, K.~J.}
\newblock \bibinfo{title}{Cavity opto-mechanics}.
\newblock \emph{\bibinfo{journal}{Opt. Express}} \textbf{\bibinfo{volume}{15}},
  \bibinfo{pages}{17172--17205}
\newblock  (\bibinfo{year}{2007}).

\bibitem{law}
\bibinfo{author}{Law, C.~K.}
\newblock \bibinfo{title}{Interaction between a moving mirror and radiation
  pressure: A hamiltonian formulation}.
\newblock \emph{\bibinfo{journal}{Phys. Rev. A}} \textbf{\bibinfo{volume}{51}},
  \bibinfo{pages}{2537--2541}
\newblock  (\bibinfo{year}{1995}).

\bibitem{gardiner}
\bibinfo{author}{Gardiner, C.~W.} \& \bibinfo{author}{Collett, M.~J.}
\newblock \bibinfo{title}{Input and output in damped quantum systems: Quantum
  stochastic differential equations and the master equation}.
\newblock \emph{\bibinfo{journal}{Phys. Rev. A}} \textbf{\bibinfo{volume}{31}},
  \bibinfo{pages}{3761--3774}
\newblock  (\bibinfo{year}{1985}).

\bibitem{Marquardt_2006}
\bibinfo{author}{Marquardt, F.}, \bibinfo{author}{Harris, J. G.~E.} \&
  \bibinfo{author}{Girvin, S.~M.}
\newblock \bibinfo{title}{Dynamical multistability induced by radiation
  pressure in high-finesse micromechanical optical cavities}.
\newblock \emph{\bibinfo{journal}{Phys. Rev. Lett.}}
  \textbf{\bibinfo{volume}{96}}, \bibinfo{pages}{103901}
\newblock  (\bibinfo{year}{2006}).

\bibitem{Zhu_2014}
\bibinfo{author}{Zhu, W.} \& \bibinfo{author}{Crozier, K.~B.}
\newblock \bibinfo{title}{Quantum mechanical limit to plasmonic enhancement as
  observed by surface-enhanced raman scattering}.
\newblock \emph{\bibinfo{journal}{Nat Commun}} \textbf{\bibinfo{volume}{5}}
\newblock  (\bibinfo{year}{2014}).

\bibitem{Ghamsari_2015}
\bibinfo{author}{Ghamsari, B.~G.}, \bibinfo{author}{Olivieri, A.},
  \bibinfo{author}{Variola, F.} \& \bibinfo{author}{Berini, P.}
\newblock \bibinfo{title}{Frequency pulling and line-shape broadening in
  graphene raman spectra by resonant stokes surface plasmon polaritons}.
\newblock \emph{\bibinfo{journal}{Phys. Rev. B}} \textbf{\bibinfo{volume}{91}},
  \bibinfo{pages}{201408}
\newblock  (\bibinfo{year}{2015}).

\bibitem{agarwal_2013}
\bibinfo{author}{Agarwal, G.~S.} \& \bibinfo{author}{Jha, S.~S.}
\newblock \bibinfo{title}{Multimode phonon cooling via three-wave parametric
  interactions with optical fields}.
\newblock \emph{\bibinfo{journal}{Phys. Rev. A}} \textbf{\bibinfo{volume}{88}},
  \bibinfo{pages}{013815}
\newblock  (\bibinfo{year}{2013}).

\bibitem{jackson}
\bibinfo{author}{Jackson, J.~D.}
\newblock \emph{\bibinfo{title}{{Classical Electrodynamics Third Edition}}}
\newblock  (\bibinfo{publisher}{Wiley}, \bibinfo{year}{1998}).

\bibitem{landau}
\bibinfo{author}{Landau, L.~D.}, \bibinfo{author}{Pitaevskii, L.~P.} \&
  \bibinfo{author}{Lifshitz, E.~M.}
\newblock \emph{\bibinfo{title}{{Electrodynamics of Continuous Media, Second
  Edition: Volume 8 (Course of Theoretical Physics)}}}
\newblock  (\bibinfo{publisher}{Butterworth-Heinemann}, \bibinfo{year}{1984}).

\bibitem{wang}
\bibinfo{author}{Wang, F.} \& \bibinfo{author}{Shen, Y.~R.}
\newblock \bibinfo{title}{General properties of local plasmons in metal
  nanostructures}.
\newblock \emph{\bibinfo{journal}{Physical Review Letters}}
  \textbf{\bibinfo{volume}{97}}
\newblock  (\bibinfo{year}{2006}).

\bibitem{wilson}
\bibinfo{author}{Wilson, E.}, \bibinfo{author}{Decius, J.} \&
  \bibinfo{author}{Cross, P.}
\newblock \emph{\bibinfo{title}{Molecular Vibrations: The Theory of Infrared
  and Raman Vibrational Spectra}}
\newblock  (\bibinfo{publisher}{Dover Publications}, \bibinfo{year}{1955}).

\bibitem{Savage}
\bibinfo{author}{Savage, K.~J.}, \bibinfo{author}{Hawkeye, M.~M.},
  \bibinfo{author}{Esteban, R.}, \bibinfo{author}{Borisov, J., Andrei
  G.and~Aizpurua} \& \bibinfo{author}{Baumberg, J.~J.}
\newblock \bibinfo{title}{Revealing the quantum regime in tunnelling
  plasmonics}.
\newblock \emph{\bibinfo{journal}{Nature}} \textbf{\bibinfo{volume}{491}}
\newblock  (\bibinfo{year}{2012}).

\bibitem{watanabe}
\bibinfo{author}{Watanabe, H.}, \bibinfo{author}{Hayazawa, N.},
  \bibinfo{author}{Inouye, Y.} \& \bibinfo{author}{Kawata, S.}
\newblock \bibinfo{title}{Dft vibrational calculations of rhodamine 6g adsorbed
  on silver analysis of tip-enhanced raman spectroscopy}.
\newblock \emph{\bibinfo{journal}{The Journal of Physical Chemistry B}}
  \textbf{\bibinfo{volume}{109}}, \bibinfo{pages}{5012--5020}
\newblock  (\bibinfo{year}{2005}).

\bibitem{humbert}
\bibinfo{author}{Humbert, C.}, \bibinfo{author}{Pluchery, O.},
  \bibinfo{author}{Lacaze, E.}, \bibinfo{author}{Tadjeddine, A.} \&
  \bibinfo{author}{Busson, B.}
\newblock \bibinfo{title}{A multiscale description of molecular adsorption on
  gold nanoparticles by nonlinear optical spectroscopy}.
\newblock \emph{\bibinfo{journal}{Phys. Chem. Chem. Phys.}}
  \textbf{\bibinfo{volume}{14}}, \bibinfo{pages}{280--289}
\newblock  (\bibinfo{year}{2012}).

\bibitem{Narula_2010}
\bibinfo{author}{Narula, R.}, \bibinfo{author}{Panknin, R.} \&
  \bibinfo{author}{Reich, S.}
\newblock \bibinfo{title}{Absolute raman matrix elements of graphene and
  graphite}.
\newblock \emph{\bibinfo{journal}{Phys. Rev. B}} \textbf{\bibinfo{volume}{82}},
  \bibinfo{pages}{045418}
\newblock  (\bibinfo{year}{2010}).

\bibitem{Zhang2013}
\bibinfo{author}{Zhang, R.} \emph{et~al.}
\newblock \bibinfo{title}{Chemical mapping of a single molecule by
  plasmon-enhanced raman scattering}.
\newblock \emph{\bibinfo{journal}{Nature}} \textbf{\bibinfo{volume}{498}},
  \bibinfo{pages}{82--6}
\newblock  (\bibinfo{year}{2013}).

\bibitem{Jiang_2015}
\bibinfo{author}{Jiang, S.} \emph{et~al.}
\newblock \bibinfo{title}{Distinguishing adjacent molecules on a surface using
  plasmon-enhanced raman scattering}.
\newblock \emph{\bibinfo{journal}{Nat Nano}} \emph{\bibinfo{volume}{advance
  online publication}}
\newblock  (\bibinfo{year}{2015}).

\end{thebibliography}
\end{document}